\documentclass[onecolumn,draftcls,12pt]{IEEEtran}
\usepackage{verbatim}
\usepackage{amsfonts}
\usepackage{amssymb}
\usepackage{stfloats}
\usepackage{cite}
\usepackage[colorlinks=true, urlcolor=black, linkcolor=black]{hyperref}
\usepackage{psfrag}
\usepackage{subfigure}
\usepackage{amsmath}
\usepackage{array}
\usepackage{epstopdf}
\usepackage{authblk}
\usepackage{graphicx} 
\usepackage{amsthm} 
\usepackage{lipsum}
\usepackage{verbatim} 
\usepackage{authblk}
\usepackage{mathtools}
\usepackage{cuted}
\usepackage[lined,boxed,ruled]{algorithm2e}
\usepackage{booktabs}
\usepackage{subfigure}
\usepackage{siunitx}
\usepackage{mathtools}
\usepackage{soul} 
\usepackage[]{mdframed}
\usepackage{setspace}
\usepackage{multicol}

\newcommand{\bphi}{\boldsymbol{\phi}}
\newcommand{\bPhi}{\boldsymbol{\Phi}}

\newtheorem{Lemma}{Lemma}

\newtheorem{Proposition}{Proposition}
\newtheorem{Remark}{Remark}

\title{On-the-fly Communication-and-Computing to Enable Representation Learning for Distributed Point Clouds}
\author{}

\makeatletter
\newcommand{\removelatexerror}{\let\@latex@error\@gobble}
\makeatother

\begin{document}
\author{{Xu~Chen},~{Hai~Wu}, and~{Kaibin~Huang}

\thanks{X. Chen, H. Wu, and K. Huang are with the Department of Electrical and Electronic Engineering, The University of Hong Kong, Hong Kong (Email: \{chenxu, wuhai, huangkb\}@eee.hku.hk). Corresponding author: K. Huang.}}
\maketitle
\begin{abstract}
The advent of \emph{sixth-generation} (6G) mobile networks introduces two groundbreaking capabilities: sensing and artificial intelligence (AI). Sensing leverages multi-modal sensors to capture real-time environmental data, while AI brings powerful models to the network edge, enabling intelligent \emph{Internet-of-Things} (IoT) applications. These features converge in the \emph{Integrated Sensing and Edge AI} (ISEA) paradigm, where edge devices collect and locally process sensor data before aggregating it centrally for AI tasks. Point clouds (PtClouds), generated by depth sensors, are crucial in this setup, supporting applications such as autonomous driving and mixed reality. However, the heavy computational load and communication demands of PtCloud fusion pose challenges. To address these, the FlyCom$^2$ framework is proposed, optimizing distributed PtCloud fusion through on-the-fly communication and computing, namely streaming on-sensor processing, progressive data uploading integrated communication-efficient AirComp, and the progressive output of a global PtCloud representation. FlyCom$^2$ distinguishes itself by aligning PtCloud fusion with Gaussian process regression (GPR), ensuring that global PtCloud representation progressively improves as more observations are received. Joint optimization of local observation synthesis and AirComp receiver settings is based on minimizing prediction error, balancing communication distortions, data heterogeneity, and temporal correlation.  This framework enhances PtCloud fusion by balancing local processing demands with efficient central aggregation, paving the way for advanced 6G applications. Validation on real-world datasets demonstrates the efficacy of FlyCom$^2$, highlighting its potential in next-generation mobile networks.
\end{abstract}

\section{Introduction}
Two essential capabilities, sensing and \emph{artificial intelligence} (AI), have been identified as novel features in \emph{sixth-generation} (6G) mobile networks~\cite{ITU}. 
Sensing enables 6G networks to leverage multi-modal sensors, such as radar, cameras, and LiDAR, to capture real-time information from the physical world. 
AI capability refers to the widespread deployment of powerful AI models at the network edge to support intelligent \emph{Internet-of-Things} (IoT) applications~\cite{edgeAI_Letaief,Push_AI_GX}. 
These two distinctive 6G functions are envisioned to coordinate seamlessly, automating various 6G applications, including robotic control, autonomous driving, and digital twins~\cite{Joint_Sensing_Feng}. 
This integration gives rise to the novel paradigm known as \emph{Integrated Sensing and Edge AI} (ISEA)~\cite{Aggregationgain_JSAC}.
In ISEA, edge devices act as distributed sensors, collecting sensory data from their surroundings. After preliminary local processing, this multi-sensor data is aggregated by a central server over wireless channels for central fusion and downstream AI tasks~\cite{Aggregationgain_JSAC,airpooling2023}. 
With the rapid commoditization of depth sensors (e.g., LiDAR and structured light), \emph{point clouds} (PtClouds) have become a vital sensory data modality in sensing applications~\cite{Aggregationgain_JSAC} and will find applications in various ISEA services, including scene reconstruction in autonomous driving and holographic displays in mixed reality.
A PtCloud consists of points, each with a spatial coordinate representing occupancy and an attribute reflecting luminance or color~\cite{PCanalysis}. 
Therefore, PtClouds preserve rich geometric information about the environment, enabling precise three-dimensional (3D) vision through PtCloud-based AI algorithms~\cite{DLforPC_Guo,poinconv_PC}. A key component of PtCloud-based ISEA is to perform distributed PtCloud fusion to generate a comprehensive environmental description, which however, is hindered by the heavy computational complexity of on-sensor PtCloud processing and communication bottlenecks due to high-dimensional PtCloud aggregation~\cite{PCinAir,Voca_Zhiyan}. 
For example,  modern depth sensors can generate millions of data points per second requiring processing and delivery~\cite{PCinAir}. 
These challenges motivate the current work, which proposes an \emph{on-the-fly communication-and-computing} (FlyCom$^2$) framework for efficient distributed PtCloud fusion in ISEA.

Compared to conventional sensory data, such as images, raw PtClouds exhibit significant data irregularity and heterogeneity. 
PtCloud points captured by edge sensors are often unordered and non-uniformly distributed in 3D space. This is due to the shape characteristics of the objects being represented and the limited fields of perception and distinct view angles of edge sensors~\cite{PCinAir,Voca_Zhiyan}. 
Furthermore, due to attenuation and reduced reflectivity, PtCloud points are denser near the sensor and become sparser with increasing distance~\cite{PCinAir}.
Addressing the data irregularity of PtClouds has driven advancements in distributed PtCloud fusion, which aggregates knowledge from distributed sensors to form a uniform exhaustive observation, thereby enhancing perception performance~\cite{OpenV2V,Model_Split_Li,F-cooper,Cooper}.

Extensive efforts have been undertaken to advance distributed PtCloud fusion for efficient processing, categorized into three distinct ISEA architectures: AI-on-Server ISEA, split ISEA, and AI-on-Sensor ISEA.
In AI-on-Sensor ISEA, advanced perception models, such as PointPillars~\cite{PointPillars}, are deployed directly on edge sensors. These sensors perform comprehensive PtCloud processing independently. The processed results from individual sensors are subsequently integrated using straightforward methods like summation, averaging, or majority voting (see, e.g.,~\cite{OpenV2V}).
Split ISEA, on the other hand, leverages split inference, partitioning the AI model into two sub-models, with one residing on the sensor and the other on the server~\cite{Model_Split_Li}. This architecture enables multi-view sensing fusion by aggregating intermediate features derived from on-sensor computation, thus enhancing data processing efficiency~\cite{airpooling2023,Voca_Zhiyan,F-cooper}.
The split ISEA and AI-on-Sensor ISEA suffer relatively high complexity of local calculations, making them not applicable for lightweight sensors and maintaining in-network sensing task lifetime.
To alleviate the computational burden on lightweight sensors, this architecture delegates complex data analytics to a central, high-performance server. 
Sensors are only responsible for basic pre-processing tasks and using the PtCloud feedback from sensors, the central server develops a uniform representation, such as a generative model, for the distributed PtClouds~\cite{Cooper, AENetwork_ICML2018}. This model can then predict spatial attributes for use in various PtCloud applications.


Considering distributed PtCloud fusion within the AI-on-Sensor architecture, this work aims to address an open problem in representation learning for distributed PtCloud fusion.
Representation learning for PtCloud fusion encompasses shape reconstruction and attribute rendering.
Shape reconstruction refers to re-identifying PtCloud spatial occupancy at the server, with octree-based space voxelization being a standard and popular approach in the applications (see, e.g.,~\cite{Encoding_octree}). 
In this method, the 3D space is divided into eight equally sized octants, with occupancy coded by 0-1 bits and the process continues by partitioning occupied octants into smaller ones, achieving coarse-to-fine voxelization. 
This method accounts for PtCloud sparsity, enabling efficient implementation within limited memory and generating compact bit sequences for fast occupancy feedback~\cite{Encoding_octree}.
Attribute rendering is to recover a high-quality attribute distribution on the given shape. 
Compared to shape reconstruction, attribute rendering poses significant challenges for wireless aggregation due to its larger data volume. For instance, luminance feedback requires 16 times more data, and RGB feedback requires 48 times more data under 16-bit quantization. 
To accelerate the PtCloud attribute feedback, spatial-aware PtCloud compression has been proposed (see, e.g.,~\cite{TIP_PCcompression,HoloCast_1,HoloCast_2}), which reduces raw attributes into a low-dimensional vector based on PtCloud spatial correlations. 
However, these methods require one-shot on-sensor processing with the computational complexity increasing super-linearly with the PtCloud voxel population, making them unaffordable for sensors with limited computational resources.
At the same time, delivering reduced PtCloud attributes, that still reveal a high dimensionality (e.g., 223617 to 301336 voxels per frame~\cite{TIP_PCcompression}), still imposes a communication bottleneck, especially with many sensors involved. 
These algorithms also need to wait for octree-based voxelization before attribute compression, misaligning with the incremental process of shape reconstruction.
In the current work, a FlyCom$^2$ framework is proposed to overcome the above challenges.
Its distinctive feature is to enable streaming on-sensor PtCloud processing with low complexity and progressive uploading of local observations along with the incremental shape reconstruction based on octree search.
Furthermore, the streaming communication and computation operations are executed in parallel to halve the end-to-end latency and efficient aggregation is enabled through integrating the progressive uploading and \emph{over-the-air computation} (AirComp).
AirComp addresses the multi-access problem by leveraging the superposition property of simultaneously transmitted analog waveforms to integrate aggregation computation with data uploading~\cite{Joint_Sensing_Feng,GXZhuAirComp2019}. 
Extensive research has applied AirComp in communication-efficient distributed computing and learning, such as AirComp-based federated learning in which a rich set of AirComp techniques are proposed to optimize the training efficiency~\cite{Air_distlearning,MXTao_powercontrol,Ding2020TWC}. 
Recently, AirComp has gained popularity in ISEA for enabling various sensory data aggregation functions, including averaging~\cite{Aggregationgain_JSAC,ISCC_DZ}, maximization~\cite{airpooling2023}, and spatial feature fusion~\cite{Voca_Zhiyan}.

The proposed FlyCom$^2$ framework shares some common features with relevant works.
To facilitate distributed PtCloud fusion, FlyCom$^2$ uses a 3D Gaussian process to describe the spatial distribution of PtCloud attributes, similar to the PtCloud processing in~\cite{TIP_PCcompression}.
Based on this assumption, each sensor generates a low-dimensional observation from their on-sensor PtClouds through linear projection using an observation matrix designed from the GP statistics.
Then, a \emph{single-input-multi-output} (SIMO) channel is considered to enable observation aggregation by spatially multiplexed AirComp~\cite{Aggregationgain_JSAC}, where the popular zero-forcing based transmit power control is adopted to align the signal amplitude of distributed sensors.
Despite common modeling and processing settings, existing algorithms face challenges in achieving protocol optimality in FlyCom$^2$ due to the following difficulties.
The main issue is that FlyCom$^2$ requires the progressive output of a global representation using currently received and previously accumulated observations.
Furthermore, the progressive property implies that each current design in FlyCom$^2$, including local observation synthesis, AirComp aggregation and global fusion, should take into account the effects of historical accumulations for overall fusion performance improvement.
This can not be supported by simply extending existing algorithms, that feature one-shot compression for point-to-point PtCloud feedback (e.g.,~\cite{TIP_PCcompression}) or one-shot  AirComp error minimization (e.g.,~\cite{Aggregationgain_JSAC}), to a progressive version.
At the same time, data heterogeneity and channel diversity among edge sensors necessitate a joint optimization of the communication-and-computing operations.
A sensor capturing informative PtClouds could suffer a poor channel condition.
To balance the tradeoff between channel distortion suppression and PtCloud feedback, the local observation matrices and AirComp receiver should be jointly manipulated based on an end-to-end fusion performance metric.

These difficulties are overcome in the proposed FlyCom$^2$ framework by aligning the global PtCloud fusion using progressively received observations with the process of Gaussian process regression (also known as Kriging~\cite{GPR_predictor2010}).
Gaussian process regression aims to predict the entire Gaussian process, i.e., the global PtCloud representation, using a limited set of samples, i.e., the observations progressively arriving at the server.
Exploiting the regression method further allows for the performance of PtCloud representation learning to be characterized by prediction error, guiding the end-to-end design of local observation matrices and AirComp. 
The key contributions and insights of this work are summarized as follows:
\begin{itemize}
    \item \textbf{FlyCom$^2$-Enabled Distributed PtCloud Fusion:}
     A FlyCom$^2$ framework is present for enabling efficient distributed PtCloud fusion under an AI-on-Server architecture. In FlyCom$^2$, each sensor fetches the batch of PtCloud attributes corresponding to the voxels currently generated in the octree-based PtCloud voxelization and generates a low-dimensional vector by using an optimized observation matrix. Then, the edge sensors stream the obtained observation simultaneously to the server over a SIMO channel. Once the central sensor receives the observations aggregated via AirComp, it performs a global Gaussian process regression using a closed-form predictor to learn a uniform representation for the distributed PtClouds.  
     
    \item \textbf{Communication-Computing Integrated Design:}
    Joint optimization of AirComp and observation matrix design is performed based on the end-to-end fusion performance measured by prediction error. We successively explore ideal distributed PtCloud fusion using noiseless one-shot (i.e. ideal) aggregation, AirComp-based one-shot aggregation, and FlyCom$^2$-enabled fusion. To this end, it is first demonstrated that given arbitrary observation matrices, the AirComp-based observation aggregation can be optimized by incorporating the observation matrices into the transmit power constraints in AirComp and exploiting a centroid AirComp receiver (see, e.g.,~\cite{GXZhuAirComp2019}).  Then, the observation matrix design is formulated as a ratio trace maximization problem similar to \emph{linear discriminant analysis} (LDA)~\cite{LDA}, optimally solved via generalized eigenvalue decomposition. The decomposable matrices involved in the decomposition process change across cases, reflecting different fundamental trade-offs. For ideal distributed PtCloud fusion, they are the PtCloud sample’s intra-correlation and inter-correlation matrices, simply due to the motivation of capturing more knowledge from local PtClouds for global prediction. For AirComp-based one-shot aggregation, the self-correlation matrix includes an additional term for effective channel gains, leading to the generated observation matrices balancing the AirComp distortions and data heterogeneity. These results are finally extended to the case of FlyCom$^2$, where decomposable matrices are further adjusted to account for temporal dependence in streaming operations.
    
    \item \textbf{Termination Rule for FlyCom$^2$:}
    Theoretical analysis shows that prediction error in FlyCom$^2$ decreases with continued operations, offering a flexible trade-off between fusion performance and end-to-end latency, adjustable based on task requirements. Furthermore, we show that the temporal error reduction during FlyCom$^2$ will converge to zero after sufficient accumulation of PtCloud observations, suggesting a convergence-based termination rule for the progressive operations.
    
\end{itemize}
The proposed framework and the relevant insights are validated in experiments using the popular real-world PtCloud dataset of Microsoft Voxelized Upper Bodies~\cite{TIP_PCcompression,MVUB_data}. The remainder of this paper is organized as follows. The
distributed sensing and communication models are introduced
in Section~\ref{section:model}. Then, Section~\ref{section:protocol} introduces preliminaries of Gaussian process regression and gives an overview of the proposed protocol of FlyCom$^2$-based distributed PtCloud fusion. The joint design of AirComp and observation matrices within different cases is presented in Section~\ref{section:design}, followed by performance analysis and protocol termination design in Section~\ref{section:stopping}. Experimental results are provided in Section~\ref{section:experiments}, followed by concluding remarks in Section~\ref{section:conclusions}.

%

\begin{figure*}[t]
    \centering
    \includegraphics[width=0.9\textwidth]{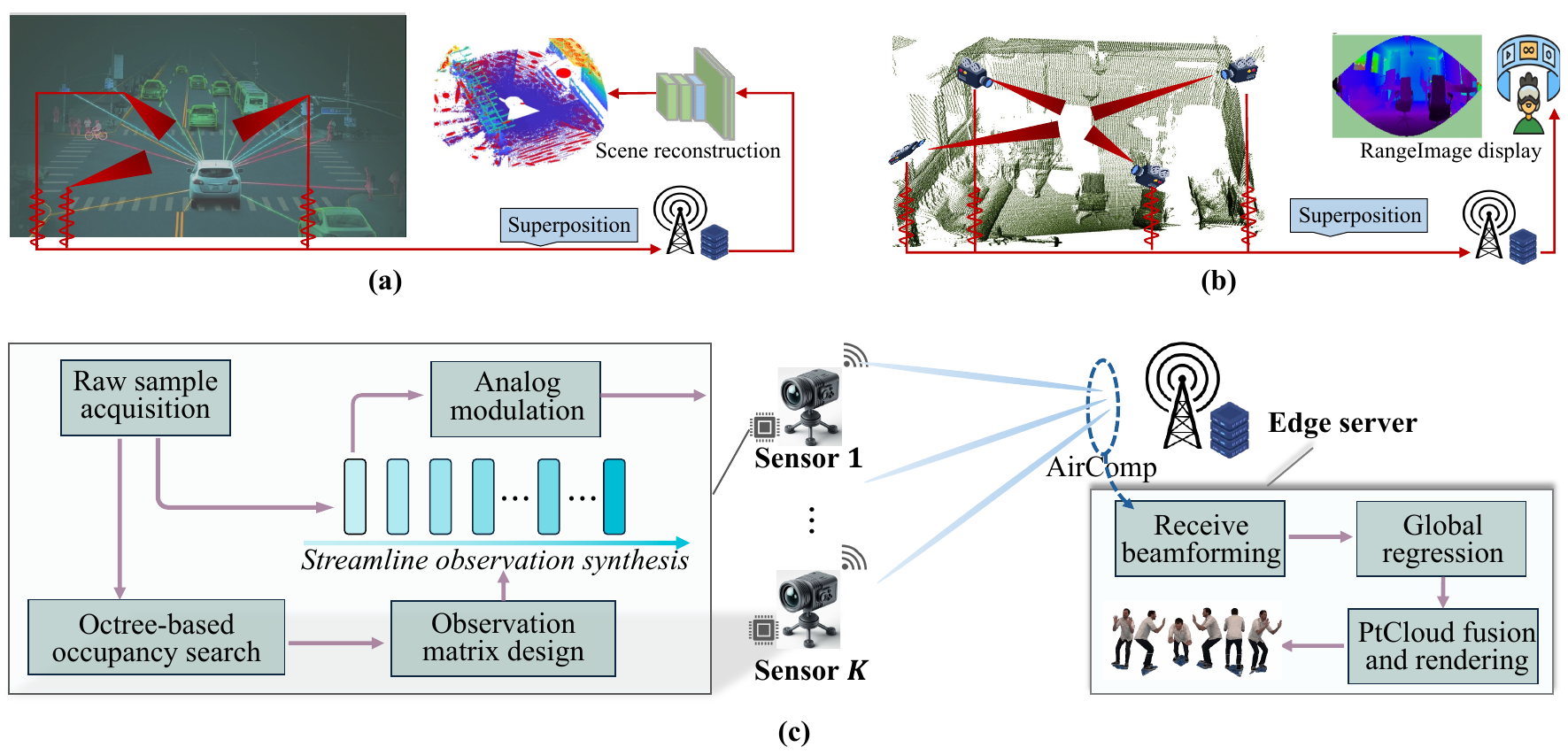}
    \caption{An ISEA system for environmental perception in the context of (a) autonomous driving [\href{https://researchleap.com/research-in-autonomous-driving-a-historic-bibliometric-view-of-the-research-development-in-autonomous-driving/}{Figure source}] and (b) mixed reality [\href{https://ieeexplore-ieee-org.eproxy.lib.hku.hk/stamp/stamp.jsp?tp=&arnumber=5980567}{Figure source}], supported by (c) the proposed FlyCom$^2$ framework.}
    \label{fig:system_model}
\end{figure*}
\section{System Model and Metrics}\label{section:model}
We consider the support of distributed PtCloud fusion in an ISEA system, as illustrated in Fig.~\ref{fig:system_model}. The relevant models, operations, and metrics are described as follows.
\subsection{Distributed PtCloud Sensing}
Consider distributed sensing, where $K$ depth cameras are employed to sense the environment and their captured data points will be collected by the server for downstream tasks. A point cloud represents a collection of points, where each point in PtCloud comprises its spatial positions as well as the corresponding real-valued attributes, such as luminance and colors. The raw PtClouds will be transformed into a formatted representation via voxelization to facilitate subsequent processing and task execution. In voxelization, the points are grouped into a collection of regular grids, called voxels, according to the spatial association relations. Then, one point is taken from each voxel to approximate all the points in that voxel~\cite{PCanalysis}. At an arbitrary sensor, say device $k$, let the voxel-represented PtCloud be denoted by $\mathcal{P}_k = \{(g_{k,n},\mathbf{s}_{k,n}\}$, where $\mathbf{s}_{k,n}\in\mathbb{R}^{3\times 1}$ represent the 3D coordinates of the $n$-th voxel and $g_{k,n}\in\mathbb{R}$ denotes the estimated scalar attribute.
\subsubsection{PtCloud Voxelization and Occupancy Feedback}
\begin{figure}[t]
    \centering
    \includegraphics[width=0.4\textwidth]{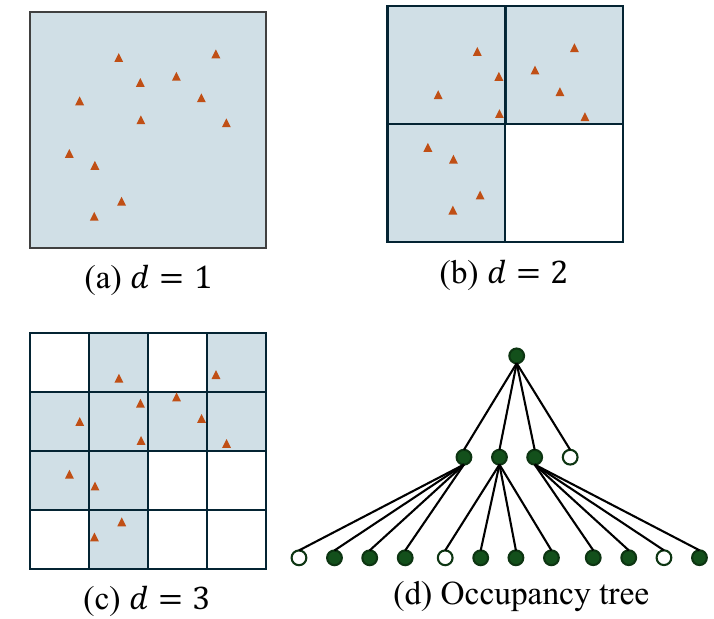}
    \caption{Illustration of octree-based search of spatial occupancy in a 2D space (quadtree), where octants in 3D space reduce into cubes and each cube will have $4$ child cubes with their occupancy being indicated by $4$ bits. In the figure, orange triangles are PtCloud points. The tree nodes are composed of filled (bit $``1"$) or unfilled (bit $``0"$) green circles indicating occupied or unoccupied cubes respectively. The generated bit sequence from depth $d=1$ to depth $d=3$ is $1\ 1110\ 0111\ 0111\ 1101$.}
    \label{fig:octree}
\end{figure}
We consider the popular octree search method to complete the space voxelization, representing compactly the point occupancy in the space into a bit sequence~\cite{octree_squeeze}. As examplified in Fig.~\ref{fig:octree}, an octree efficiently stores point clouds by dividing the input space into equally sized octants and achieving coarse-to-fine partition resolution by progressively increasing the partition depth. Then, the PtCloud occupancy is described by the octree nodes that can be encoded into compact bit sequences, as elaborated as follows. Let $\mathcal{S}_{d,j}$, $1\leq j\leq 2^{d-1}$, denote the 3D space corresponding to the $j$-th octant at depth $d$. Also, define $\mathcal{V}_i^{(d,j)}$, $1\leq i\leq 8$ be the $i$-th child octant of the octant $\mathcal{S}_{d,j}$. That is $\mathcal{S}_{d,j} = \bigcup_{1\leq i\leq 8}\mathcal{V}_i^{(d,j)}$. For the octant $\mathcal{S}_{d,j}$, a sensor with the dataset $\mathcal{P}_k$ will generate an $8$-bit sequence, denoted by $\mathbf{a}_{d,j} = [a_1^{(d,j)},a_2^{(d,j)},\cdots,a_8^{(d,j)}]$, to indicate the occupancy of its $8$ child octants: 
\begin{equation}\label{eq:occupancy}
    a_i^{(d,j)} =\begin{cases}1,&\quad \exists \mathbf{s}_{k,n}\in \mathcal{V}_i^{(d,j)},\\
    0,&\quad \mathrm{Otherwise}.
    \end{cases}
\end{equation}
The octree occupancy search only subdivides the occupied octants and then encodes their child octants' occupancy. This allows for the generated occupancy sequence to adapt to varying levels of sparsity with a compact representation for fast feedback. In this work, we assume that the occupancy bit sequences progressively generated by each device are fed back to the server over dedicated channels (see, e.g.,~\cite{Encoding_octree}).

\subsubsection{Distributed PtCloud Fusion via Representation Learning}
Given the spatial identification of PtClouds, the distributed PtCloud fusion is achieved by learning a unified attribute representation from local PtClouds. Without loss of generity, in a PtCloud $\mathcal{P}_k = \{(g_{k,n},\mathbf{s}_{k,n}\}$, the attribute value of a point relate to its spatial position as:
\begin{equation}\label{eq:attribute_model}
    g_{k,n} = f(\mathbf{s}_{k,n}) + \epsilon_{k,n},\quad \mathbf{s}_{k,n}\in\mathcal{C},
\end{equation}
where $f(\mathbf{s}_{k,n})$ denotes the true attribute value corresponding to the position $\mathbf{s}_{k,n}$, $\epsilon_{k,n}$ represents the estimation error, and $\mathcal{C}$ denotes the domain of $f(\cdot)$. $\epsilon_{k,n}$ is modeled as an \emph{independent and identically distributed} (i.i.d.) Gaussian variable $\mathcal{N}(0,\sigma_{\sf e}^2)$ and the variance $\sigma_{\sf e}^2\rightarrow 0$ as the resolution of the voxelization is sufficient large~\cite{TIP_PCcompression}. The function $f(\cdot)$ is the generative representation for the PtClouds, characterizing the spatial distributions of PtCloud attributes~\cite{PCinAir}. The goal of distributed PtCloud fusion is to estimate the function $f(\cdot)$ using local PtCloud samples~\cite{PCinAir}. Therefore, given $\tilde{f}(\cdot)$ as an estimated generative representation, the geometry information of the environment can be inferred from  $\tilde{f}(\cdot)$ by simply resampling:
\begin{equation}\label{eq:retrival}
    \mathsf{attribute\ of\ position\ }\mathbf{s} = \tilde{f}(\mathbf{s}), \quad \mathbf{s}\in\mathcal{C}.
\end{equation}

The problem of identifying the form of $f(\cdot)$ is addressed by the FlyCom$^2$ protocol proposed in Section~\ref{section:protocol}. It models $f(\cdot)$ as a realization of a Gaussian process similar to~\cite{TIP_PCcompression} and estimates $f(\cdot)$ within the progressive arrivals of local PtCloud observations using the technique of Gaussian process regression.

\subsection{Communication Modelling}
We consider AirComp for data aggregation in the ISEA system in Fig.~\ref{fig:system_model}. The assumptions and operations of the schemes are described as follows. The server and sensors are equipped with $M$-element array and a single antenna, respectively. Assuming a frequency non-selective channel, time is split into symbol intervals and each interval is used for transmitting one symbol. Block fading is considered such that the channel remains unchanged over a \emph{communication slot} comprising $I$ symbol intervals. Symbol-level synchronization is assumed over all sensors.

During an arbitrary communication slot, say slot $t$, the distributed sensors simultaneously transmit their linear analog modulated data symbols, $\mathbf{x}_{k,t}=[x_{k,1}^{(t)},\cdots,x_{k,I}^{(t)}]^{\top}$. This leads to the server receiving a symbol matrix:
\begin{equation}\label{eq:wireless_model}
    \mathbf{Y}_t = \sum_{k}\rho_{k,t}\mathbf{h}_{k,t}\mathbf{x}_{k,t}^{\top}+\mathbf{Z}_t,
\end{equation}
where $\rho_{k,t}$ represents transmit power, $\mathbf{h}_{k,t}$ denotes the channel vector of sensor $k$, and $\mathbf{Z}_t$ models additive channel noise with i.i.d. $\mathcal{CN}(0,\sigma^2)$ elements. 
Assuming Rayleigh fading, $\mathbf{h}_{k,t}\sim\mathcal{CN}(\mathbf{0},\mathbf{I}_M)$ and is independent between sensors. Let $\nu_{k,t}^2\overset{\triangle}{=}\mathsf{E}\left[\Vert\mathbf{x}_{k,t}\Vert^2\right]/T$ be the variance of transmitted symbols over a channel coherence block. Each sensor is constrained by a power budget of $P$, i.e., $\rho_{k,t}^2\nu_{k,t}^2\leq P$. Then, the transmit SNR is defined as $\gamma = \frac{P}{\sigma^2}$.  

Following the AirComp literature, zero-forcing transmit power control is adopted to overcome channel distortion, $\rho_{k,t} = (\mathbf{b}_t^H\mathbf{h}_{k,t})^{-1}$ with $\mathbf{b}_t\in \mathbb{C}^{N}$ being a receive combiner (see, e.g.,~\cite{GXZhuAirComp2019}). As a result, a noisy version of the average of $\{\mathbf{x}_k\}$ can be obtained as
\begin{equation}\label{eq:AirComp_avg}
    \tilde{\mathbf{x}}_t=\frac{1}{K}\mathbf{Y}_t^H\mathbf{b}_t=\frac{1}{K}\sum_{k}\mathbf{x}_{k,t} + \bar{\mathbf{z}}_t,
\end{equation}
where $\bar{\mathbf{z}}_t \overset{\triangle}{=}\mathbf{Z}_t^H\mathbf{b}_t$ denotes the caused channel distortion. Using the property of Gaussian distributions, there is $\bar{\mathbf{z}}_t\sim\mathcal{CN}(\mathbf{0},\sigma^2\Vert\mathbf{b}_t\Vert^2\mathbf{I}_I)$.
\subsection{Performance Metrics}
The estimation accuracy of reconstructing $f(\cdot)$ can be reflected by the expected deviation between the predicted values of all possible points in the desired 3D space $\mathcal{C}$ and their ground truth. Commonly, the prediction accuracy is measured by the \emph{mean squared error} (MSE) defined as~\cite{GPR_predictor2010}
\begin{align}\label{eq:MSPE}
        \mathsf{error}(f,\tilde{f})& = \frac{1}{|\mathcal{C}|}\mathsf{E}\left[\sum_{\mathbf{s}\in\mathcal{C}}\left(f(\mathbf{s})-\tilde{f}(\mathbf{s})\right)^2\right].
\end{align}


\section{FlyCom$^2$-Based Distributed PtCloud Fusion}\label{section:protocol}
\begin{figure*}[t]
    \centering
    \includegraphics[width=0.95\textwidth]{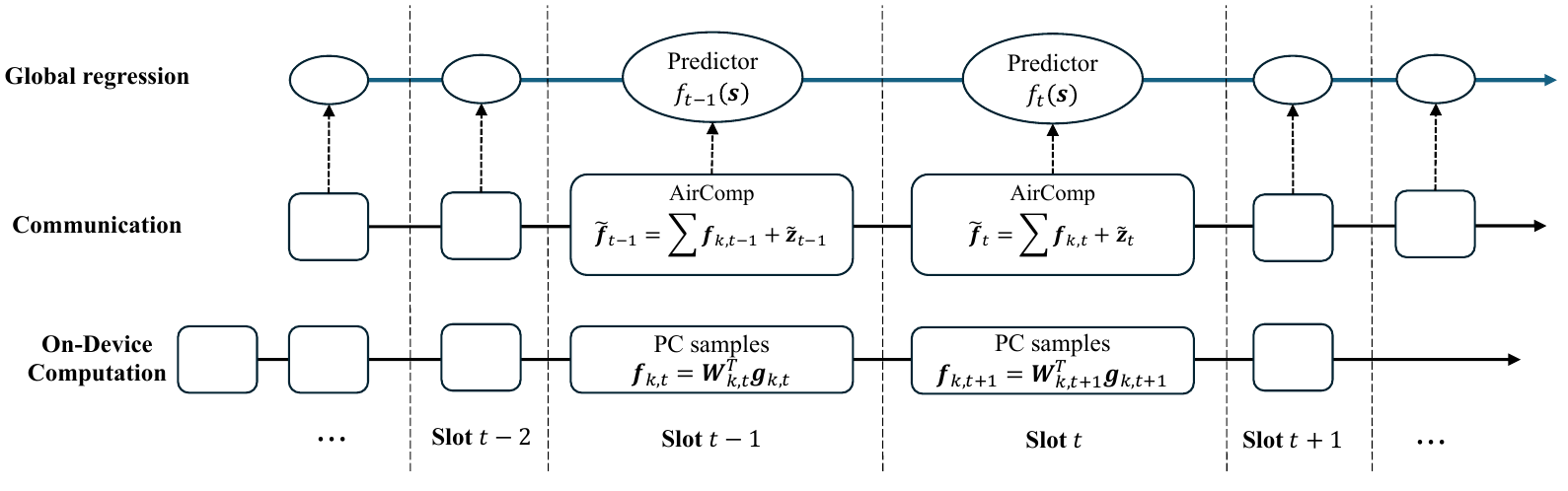}
    \caption{Factor graph of FlyCom$^2$-based distributed PtCloud fusion featuring parallel streams of on-device observation synthesis, AirComp-based aggregation, and global Gaussian process regression.}
    \label{fig:workflow}
\end{figure*}

To develop distributed PtCloud fusion over edge devices with limited computation power and communication resources, we propose a FlyCom$^2$ framework as illustrated in Fig.~\ref{fig:workflow}. In the sequel,
we first briefly introduce Gaussian process regression exploited in the proposed framework and then elaborate on the detailed FlyCom$^2$ procedures.
\subsection{Preliminaries of Gaussian Process Regression}
In this subsection, the preliminaries of Gaussian process regression are briefly summarized. Gaussian process regression is a Bayesian approach that is adept for learning the function $f(\mathbf{s})$ under a set of observations  $\{g_n=f(\mathbf{s}_n) + \epsilon_n\}$, $1\leq n\leq N$ with the observation noise $\epsilon_n\sim \mathcal{N}(0,\sigma_{\mathsf{e}}^2)$~\cite{GP_for_ML2006}. Its core idea is to treat $f(\cdot)$ as a realization of a stationary \emph{Gaussian process} $\mathcal{Z}$ due to the latter's efficient approximation of arbitrary measurable functions. The Gaussian process, $\mathcal{Z}\sim\mathcal{GP}(\mu(\mathbf{s}),\phi(\mathbf{s},\mathbf{s}^{\prime}))$, is identified by a mean function $\mu(\mathbf{s})$  and a covariance function (i.e. kernel), $\phi(\mathbf{s},\mathbf{s}^{\prime})$, that describes the correlation between $f(\mathbf{s})$ and $f(\mathbf{s}^{\prime})$. Without loss of generality, the mean function is commonly set as $\mu(\mathbf{s}) = 0$ via normalization. 

The Gaussian process regression is to learn the function $f(\cdot)$ using a set of available samples $\{(g_n,\mathbf{s}_n)\}$ and the prior statistic knowledge of the kernels $\phi(\mathbf{s},\mathbf{s}^{\prime})$. In the literature, popular kernels include the squared exponential kernel $\phi(\mathbf{s},\mathbf{s}^{\prime}) = \beta\exp\left(-\frac{\Vert\mathbf{s}-\mathbf{s}^{\prime}\Vert^2}{2\theta^2}\right)$ and the Ornstein-Uhlenbeck kernel $\phi(\mathbf{s},\mathbf{s}^{\prime}) = \beta\theta^{\Vert\mathbf{s}-\mathbf{s}^{\prime}\Vert}$, where $(\beta,\theta)$ denotes the hyperparameter. These kernels are motivated by the intuition that the outputs $f(\mathbf{s})$ and $f(\mathbf{s}^{\prime})$ are expected to be similar if $\mathbf{s}$ is close to $\mathbf{s}^{\prime}$. Furthermore, the variance of additional observation noise, $\sigma_{\mathsf{e}}^2$, can be not available in prior. To obtain these hyperparameter $\Theta = (\beta,\theta,\sigma_{\mathsf{e}})$, one can solve the maximization of the log marginal likelihood over a given training dataset~\cite{GP_for_ML2006}.

Since $f(\cdot)$ is a realization of $\mathcal{GP}(0,\phi(\mathbf{s},\mathbf{s}^{\prime}))$ and the observation noise also follows an independent Gaussian distribution, the value of $f(\mathbf{s})$ for an arbitrary input $\mathbf{s}$ and existing training data $\mathbf{g} = [g_1,g_2,\cdots,g_N]^{\top}$ shall have a joint distribution expressed as
\begin{equation}\label{eq:GPRjoint_dist}
    \begin{bmatrix}
        \mathbf{g}\\
        f(\mathbf{s})
    \end{bmatrix}\sim\mathcal{N}\left(        \begin{bmatrix}
        \mathbf{0}\\
        0
    \end{bmatrix},        \begin{bmatrix}
        \bPhi+\sigma_{\mathsf{e}}^2\mathbf{I}&\bphi(\mathbf{s})\\
        \bphi^{\top}(\mathbf{s})&\phi(\mathbf{s},\mathbf{s})
    \end{bmatrix}\right),
\end{equation}
where $\bphi(\mathbf{s}) = [\phi(\mathbf{s},\mathbf{s}_1),\phi(\mathbf{s},\mathbf{s}_2),\cdots,\phi(\mathbf{s},\mathbf{s}_N)]^{\top}$ and $\bPhi\in\mathbb{R}^{N\times N}$ with its $(n,n^{\prime})$-th element computed as $\phi(\mathbf{s}_n,\mathbf{s}_{n^{\prime}})$. Therefore, one can predict the entire process $f(\cdot)$ using given observations $\{(g_n,\mathbf{s}_n)\}$ and the joint distribution~\eqref{eq:GPRjoint_dist}. The predictive distribution of $f(\mathbf{s})$, namely the conditional distribution $\mathrm{Pr}(f(\mathbf{s})|\mathbf{g},\{\mathbf{s}_n\},\mathbf{s})$, is a Gaussian distribution with mean and variance given as 
\begin{equation}\label{eq:GPR_mean}
    \mathsf{E}[f(\mathbf{s})] = \bphi^{\top}(\mathbf{s})\left(\bPhi+\sigma_{\mathsf{e}}^2\mathbf{I}\right)^{-1}\mathbf{g},
\end{equation}
\begin{equation}\label{eq:GPR_var}
    \mathsf{Var}[f(\mathbf{s})] = \phi(\mathbf{s},\mathbf{s})-\bphi^{\top}(\mathbf{s})\left(\bPhi+\sigma_{\mathsf{e}}^2\mathbf{I}\right)^{-1}\bphi(\mathbf{s}).
\end{equation}
Furthermore, an important conclusion has been obtained as shown below.
\begin{Lemma}[MSE~\cite{GPR_predictor2010}]\label{Lemma:estimator}
    \emph{The mean function in~\eqref{eq:GPR_mean} is the best linear unbiased predictor of $f(\mathbf{s})$ that achieves the minimal MSE that equals to $\mathsf{Var}[f(\mathbf{s})]$ in~\eqref{eq:GPR_var}.}
\end{Lemma}

In this work, we assume the spatial distributes of the PtCloud attributes to follow a Gaussian process. Such modeling has proved efficient in the field of PtCloud compression~\cite{TIP_PCcompression}. Then, exploiting the Gaussian process regression approach, we propose the following FlyCom$^2$ framework to learn a generative representation $f(\cdot)$ for the distributed PtCloud fusion.
\subsection{Overview of the Proposed FlyCom$^2$ Protocol}
As shown in Fig.~\ref{fig:workflow}, the proposed FlyCom$^2$ framework is to learn $f(\cdot)$ based on a stream of PtCloud samples generated at edge sensors along with the octree space occupancy. The progressive property gives the name of FlyCom$^2$, as opposed to the traditional one-shot framework requires the execution of high-dimensional data on-sensor processing and uploading~\cite{TIP_PCcompression,HoloCast_1,HoloCast_2}. The details are provided below.

\subsubsection{On-the-Fly Observation Synthesis at Sensors}
In each communication slot, each sensor computes a local observation using its on-sensor PtClouds and transmits the observation generated in the previous slot to the server in parallel. The observations are constructed during real-time detection of PtClouds one by one using globally designed observation matrices. 
Consider the progressive observation synthesis at sensor $k$ at the slot $t$. For ease of notation, we define, $\mathcal{K}_{k,t} = \{\mathcal{S}_{d,j}\}$ and $\{(g_{k,\pi(d,j)},\mathbf{s}_{k,\pi(d,j)})\}_{\mathcal{S}_{d,j}\in \mathcal{K}_{k,t}}$ respectively, as the set of the detected octants and the tuples of their corresponding coordinates and attributes. The attributes of these octants are aggregated into a vector $\mathbf{g}_{k,t} = [g_{k,\pi(d,j)}]_{\mathcal{S}_{d,j}\in\mathcal{K}_{k,t}}$. Then, sensor $k$ computes an observation from $\mathbf{g}_{k,t}$ by exploiting a simple linear projection~\cite{TIP_PCcompression,HoloCast_1,HoloCast_2}:
\begin{equation}\label{eq:local_estimate}
    \mathbf{f}_{k,t} = \mathbf{W}_{k,t}^{\top}\mathbf{g}_{k,t},
\end{equation}
where $\mathbf{\mathbf{W}}_{k,t} \in\mathbb{R}^{I\times |\mathcal{K}_{k,t}|}$ represents a real observation matrix that maps the raw PtCloud attributes onto a low-dimensional space\footnote{It deserves mentioning that $\mathbf{\mathbf{W}}_{k,t}$ could be a complex matrix to capture knowledge using doubled dimensionality, which however, can be straightforwardly extended from the current case by leveraging the concatenation of real and imaginary parts of a complex $\mathbf{f}_{k,t}$. It does not offer new insights but complicates the notations, motivating us to focus on the real observation matrix.}. The advantages of the method~\eqref{eq:local_estimate} are threefold. First, the low-dimensional projection removes the data redundancy for efficient PtCloud feedback. Second, the detected octant sets $\mathcal{K}_{k,t}$ can be different between sensors due to their distinct view angles and relative positions, leading to the issues of spatial alignment and data heterogeneity. The approach~\eqref{eq:local_estimate} allows for efficient implementation of AirComp into observation aggregation, avoiding octantwise spatial alignment between sensors and making it tractable to address the data heterogeneity through the global design of $\{\mathbf{W}_{k,t}\}$. Third, the computation of~\eqref{eq:local_estimate} serves as pre-processing with a slight computation load imposed on sensors, allowing them to be easily implemented into lightweight devices.

\subsubsection{AirComp-Based Observation Aggregation}
The local observations simultaneously streamed by sensors will be aggregated at the server via SIMO AirComp. Starting with the noiseless aggregation, a global observation, denoted by $\bar{\mathbf{f}}_t$, can be simply obtained as the summation of the local observations generated in~\eqref{eq:local_estimate} since
\begin{align}
       \bar{\mathbf{f}}_t &= \sum_{k}\mathbf{f}_{k,t} = [\mathbf{W}_{1,t}^{\top},\ \cdots,\mathbf{W}_{K,t}^{\top}]\begin{bmatrix}
       \mathbf{g}_{1,t}\\
       \vdots\\
       \mathbf{g}_{K,t}
   \end{bmatrix}\nonumber\\
   & = \bar{\mathbf{W}}_t^{\top}\bar{\mathbf{g}}_t,\label{eq:aggregation}
\end{align}
where $\bar{\mathbf{W}}_t \overset{\triangle}{=} [\mathbf{W}_{1,t}^{\top},\ \cdots,\mathbf{W}_{K,t}^{\top}]^{\top}$ denotes a global observation matrix and $\bar{\mathbf{g}}_t\overset{\triangle}{=}  [\mathbf{g}_{1,t}^{\top},\cdots,\mathbf{g}_{K,t}^{\top}]^{\top}$ is the concatenation of all data samples locally detected in the current slot. $\bar{\mathbf{f}}_t$ can be then treated as the low-dimensional projection capturing principal components of $\bar{\mathbf{g}}_t$. With a proper design of $\bar{\mathbf{W}}_t$, key knowledge in $\bar{\mathbf{g}}_t$ can be retained for guaranteeing the performance of global prediction while reducing the dimensionality of required aggregations. The operation in~\eqref{eq:aggregation} can be efficiently realized by using AirComp in~\eqref{eq:AirComp_avg}, at the expense of additional channel distortions. Specifically, the AirComped aggregation of~\eqref{eq:aggregation} will lead to the server receiving a vector symbol expressible as $\sum_{k}\mathbf{f}_{k,t} + \bar{\mathbf{z}}_t$, whose the real part gives an effective observation since the real-valued attributes. Let define $\tilde{\mathbf{f}}_t$ as the received noise version of $\bar{\mathbf{f}}_t$. Then, there is
\begin{equation}\label{eq:AirComp_agg}
    \tilde{\mathbf{f}}_t = \Re\left\{\sum_{k}\mathbf{f}_{k,t} + \tilde{\mathbf{z}}_t\right\}=\bar{\mathbf{W}}_t^{\top}\bar{\mathbf{g}}_t + \Re\left\{\bar{\mathbf{z}}_t\right\},
\end{equation}
where  $\Re\left\{\tilde{\mathbf{z}}_t\right\}$ denotes the real part of the effective AirComp distortion. According to~\eqref{eq:AirComp_avg} and the properties of complex Gaussian distributions, $\Re\left\{\bar{\mathbf{z}}_t\right\}$ is composed of i.i.d. $\mathcal{N}(0,\frac{1}{2}\sigma^2\Vert\mathbf{b}_t\Vert^2)$ entries.
\subsubsection{On-the-Fly Global Regression at a Server}

In communication slot $t$, global regression is performed using previous and currently received observations, $\{\tilde{\mathbf{f}}_{\ell}\}_{\ell<t}$ to generate a predictor for the estimation of $f(\mathbf{s})$ for $\mathbf{s}\in\mathcal{C}_t$. For ease of notation, we define $\tilde{\mathbf{f}}_{<t}$ as the aggregations of all previously received observations and replace $t$ by $<t$ in the involved symbols defined before. Under the Gaussian process modeling, the optimal predictive distribution for the distributed PtClouds, $f(\mathbf{s})$, can be estimated using the property of Gaussian process regression~\eqref{eq:GPRjoint_dist}. Specifically, both~\eqref{eq:local_estimate} and~\eqref{eq:aggregation} are linear combinations over zero-mean Gaussian variables, suggesting that $\tilde{\mathbf{f}}_{<t}$ and $\tilde{\mathbf{f}}_t$ still follow a zero-mean Gaussian distribution. Hence, the joint distribution of $f(\mathbf{s})$, $\tilde{\mathbf{f}}_{<t}$, and $\mathbf{f}_t$ can be written as
\begin{equation}\label{eq:joint_dist}
    \begin{bmatrix}
        \tilde{\mathbf{f}}_{<t}\\
        \tilde{\mathbf{f}}_t\\
        f(\mathbf{s})
    \end{bmatrix}\sim\mathcal{N}\left(    \mathbf{0},\begin{bmatrix}
    \mathbf{R}_t&\mathbf{r}_t\\
        \mathbf{r}_t^{\top} & \beta
    \end{bmatrix}\right),
\end{equation}
where $\beta$ denotes the self-covariance of an arbitrary sampling point of the Gaussian process, i.e. $\beta = \mathsf{E}[f(\mathbf{s})f(\mathbf{s})]$. The other covariance variables in~\eqref{eq:joint_dist} are given as
\begin{align}
        \mathbf{r}_t =    \begin{bmatrix}
        \mathsf{E}[\tilde{\mathbf{f}}_{<t}f(\mathbf{s})]\\
        \mathsf{E}[\tilde{\mathbf{f}}_tf(\mathbf{s})]
    \end{bmatrix} = \begin{bmatrix}
        \bar{\mathbf{W}}_{<t}^{\top}&\\
        &\bar{\mathbf{W}}_t^{\top}
    \end{bmatrix}\begin{bmatrix}
        \bphi_{<t}(\mathbf{s})\\
        \bphi_t(\mathbf{s})
    \end{bmatrix},
\end{align}
\begin{align}
    \mathbf{R}_t &= \begin{bmatrix}
\mathsf{E}[\tilde{\mathbf{f}}_{<t}\tilde{\mathbf{f}}_{<t}^{\top}]&\mathsf{E}[\tilde{\mathbf{f}}_{<t}\tilde{\mathbf{f}}_t^{\top}]\\
    \mathsf{E}[\tilde{\mathbf{f}}_t\tilde{\mathbf{f}}_{<t}^{\top}] & \mathsf{E}[\tilde{\mathbf{f}}_t\tilde{\mathbf{f}}_t^{\top}]
\end{bmatrix}\nonumber\\
&=\begin{bmatrix}
    \bar{\mathbf{W}}_{<t}^{\top}&\\
    &\bar{\mathbf{W}}_t^{\top}
\end{bmatrix}\begin{bmatrix}
    \bar{\bPhi}_{<t,<t}&\bar{\bPhi}_{<t,t}\\
    \bar{\bPhi}_{<t,t}^{\top}&\bar{\bPhi}_{t,t}
\end{bmatrix}\begin{bmatrix}
    \bar{\mathbf{W}}_{<t}&\\
    &\bar{\mathbf{W}}_t
\end{bmatrix}\nonumber\\
&\quad+ \begin{bmatrix}
    \mathbf{\Psi}_{<t}&\\
    &\frac{1}{2}\sigma^2\Vert\mathbf{b}_t\Vert^2\mathbf{I}_N
\end{bmatrix},\label{eq:overall_cov}
\end{align}
where we define
\begin{align}
    &\bphi_{<t}(\mathbf{s}) = \mathsf{E}[\bar{\mathbf{g}}_{<t}f(\mathbf{s})],\ \bphi_t(\mathbf{s}) = \mathsf{E}[\bar{\mathbf{g}}_tf(\mathbf{s})],\\
    &\bar{\bPhi}_{<t,<t}=\mathsf{E}[\bar{\mathbf{g}}_{<t}\bar{\mathbf{g}}_{<t}^{\top}],\ \bar{\bPhi}_{<t,t}=\mathsf{E}[\bar{\mathbf{g}}_{<t}\bar{\mathbf{g}}_t^{\top}],\\
    &\bar{\bPhi}_{t,t}=\mathsf{E}[\bar{\mathbf{g}}_t\bar{\mathbf{g}}_t^{\top}].
\end{align}
and $\mathbf{\Psi}_{<t}$ is a diagonal matrix with its diagonal elements representing AirComp noise power in previous slots. The computation of $\mathbf{R}_t$ and $\mathbf{r}_t$ can be computed using a given Gaussian process kernel and the position knowledge obtained during the octree occupancy search~\eqref{eq:occupancy} and feedback. Then, using Lemma~\ref{Lemma:estimator}, the optimal linear predictor for $f(\mathbf{s})$ at the current communication $t$ is given as
\begin{equation}\label{eq:global_predictor}
\boxed{f_t(\mathbf{s}) = \mathbf{r}_t^{\top}\mathbf{R}_t^{-1}\begin{bmatrix}
        \tilde{\mathbf{f}}_{<t}\\
        \tilde{\mathbf{f}}_t
    \end{bmatrix}}.
\end{equation}
 
\subsection{Discussions on Computation Complexity}
\subsubsection{Complexity at Devices}
At devices, the main computation complexity comes from the local observation systhesis~\eqref{eq:local_estimate}, where given the observation matrix with the size of $I\times|\mathcal{K}_{k,t}|$, the resulting matrix-vector complexity is $\mathcal{O}(I|\mathcal{K}_{k,t}|)$.
\begin{figure*}[!b]
 \normalsize 
 \hrulefill
    \begin{equation}\label{eq:progressive}
        \mathbf{R}_t^{-1} =    \begin{bmatrix}
        \mathbf{R}_{t-1}^{-1}+ \mathbf{R}_{t-1}^{-1}\bar{\mathbf{W}}_{<t}^{\top}\bar{\bPhi}_{<t,t}\bar{\mathbf{W}}_t\mathbf{A}^{-1}\bar{\mathbf{W}}_t^{\top}\bar{\bPhi}_{<t,t}^{\top}\bar{\mathbf{W}}_{<t}\mathbf{R}_{t-1}^{-1}  &-\mathbf{R}_{t-1}^{-1}\bar{\mathbf{W}}_{<t}^{\top}\bar{\bPhi}_{<t,t}\bar{\mathbf{W}}_t\mathbf{A}^{-1}\\
        -\mathbf{A}^{-1}\bar{\mathbf{W}}_t^{\top}\bar{\bPhi}_{<t,t}^{\top}\bar{\mathbf{W}}_{<t}\mathbf{R}_{t-1}^{-1}& \mathbf{A}^{-1} 
    \end{bmatrix}, 
    \end{equation}
        \begin{equation}\label{eq:A}
        \mathbf{A} =  \bar{\mathbf{W}}_{t}^{\top}\bar{\bPhi}_{t,t}^{\top}\bar{\mathbf{W}}_{t}+\frac{1}{2}\sigma^2\Vert\mathbf{b}_t\Vert^2\mathbf{I}_N - \bar{\mathbf{W}}_t^{\top}\bar{\bPhi}_{<t,t}^{\top}\bar{\mathbf{W}}_{<t}\mathbf{R}_{t-1}^{-1}\bar{\mathbf{W}}_{<t}^{\top}\bar{\bPhi}_{<t,t}\bar{\mathbf{W}}_t.
    \end{equation}
\end{figure*}
\subsubsection{Complexity at The Server}
The computation complexity at the server is due to aspects. First, the server is required to compute the observation matrix, which is shown in Section~\ref{section:design} to be consistent with generalized eigenvalue decomposition~\cite{GED_complexity}. In time slot $t$, consider $|\mathcal{K}_{k,t}|$ on-device samples at sensor $k$, and $I_t$ principal components (i.e., the dimensionality of the observation matrix $\bar{\mathbf{W}}_t$). Then, the computation complexity resultant from observation matrix design is $\mathcal{O}\left(I_t\left(\sum_{k=1}^K|\mathcal{K}_{k,t}|\right)^2\right)$. On the other hand, the server needs to generate the predictor~\eqref{eq:global_predictor} whose computation complexity is dominated by the matrix inverse $\mathbf{R}_t^{-1}$. The dimensionality of the accumulated observations at slot $t$ is given as $\sum_{\ell=1}^tI_{\ell}$, leading to the complexity of $\mathcal{O}\left((\sum_{\ell=1}^tI_{\ell})^3)\right)$. Such complexity can be further reduced by computing $\mathbf{R}_t^{-1}$ through $\mathbf{R}_{t-1}^{-1}$ based on~\eqref{eq:progressive} and~\eqref{eq:A}, if $\mathbf{A}$ in~\eqref{eq:A} is non-singular. This reduces the complexity to $\mathcal{O}\left(I_t(\sum_{\ell=1}^{t-1}I_{\ell})^2)\right)$, leading to the overall complexity of $\mathcal{O}\left(I_t(\sum_{k=1}^K|\mathcal{K}_{k,t}|)^2+I_t(\sum_{\ell=1}^{t-1}I_{\ell})^2)\right)$. 

\subsection{System Optimization}
Improving the performance of distributed PtCloud fusion requires optimizing the prediction accuracy of~\eqref{eq:global_predictor}, which is determined by both the observation matrix ($\bar{\mathbf{W}}_t$) design and AirComp receiver ($\mathbf{b}_t$) optimization. Specifically, it follows from Lemma~\ref{Lemma:estimator} that given the accurate computation of kernel values, $f_t(\mathbf{s})$ gives an unbiased estimation of $f(\mathbf{s})$, namely $\mathsf{E}[f_t(\mathbf{s})] = \mathsf{E}[f(\mathbf{s})]$, $\forall \mathbf{s}\in\mathcal{C}$, and the resulting variance between $f_t(\mathbf{s})$ and $f(\mathbf{s})$ is given as
\begin{align}\label{eq:variance}
    \mathsf{Var}[f_t(\mathbf{s})] \!=\! \beta\! -\! \begin{bmatrix}
        \bar{\mathbf{W}}_{<t}^{\top}\bphi_{<t}(\mathbf{s})\\
        \bar{\mathbf{W}}_t^{\top}\bphi_t(\mathbf{s})
    \end{bmatrix}^{\top}\mathbf{R}_t^{-1}\begin{bmatrix}
        \bar{\mathbf{W}}_{<t}^{\top}\bphi_{<t}(\mathbf{s})\\
        \bar{\mathbf{W}}_t^{\top}\bphi_t(\mathbf{s})
    \end{bmatrix}.
\end{align}
It then follows from~\eqref{eq:variance} that the MSE achieved by the slot-$t$ predictor in~\eqref{eq:global_predictor}, is given as
\begin{align}\label{eq:temporal_error}
    \mathsf{error}_t &= \frac{1}{|\mathcal{C}_t|}\sum_{\mathbf{s}\in\mathcal{C}_t} \mathsf{E}[(f_t(\mathbf{s})-f(\mathbf{s}))^2]\nonumber\\
    &= \beta - \mathsf{Tr}\left(\mathbf{R}_t^{-1}\mathbf{J}_t\right),
\end{align}
where $\mathcal{C}_t$ denotes the set of all occupied octants detected during the octree space search so far and for ease of notation, we define
\begin{align}
    &\mathbf{J}_t = \begin{bmatrix}
        \bar{\mathbf{W}}_{<t}^{\top}&\\
        &\bar{\mathbf{W}}_t^{\top}
    \end{bmatrix}\begin{bmatrix}
        \mathbf{Q}_{<t,<t}&\mathbf{Q}_{<t,t}\\
        \mathbf{Q}_{<t,t}^{\top}&\mathbf{Q}_{t,t}
    \end{bmatrix}\begin{bmatrix}
        \bar{\mathbf{W}}_{<t}&\\
        &\bar{\mathbf{W}}_t
    \end{bmatrix},\label{eq:overall_pre}\\
        &\mathbf{Q}_{<t,<t} = \sum_{\mathbf{s}\in\mathcal{C}_t}\bphi_{<t}(\mathbf{s})\bphi_{<t}^{\top}(\mathbf{s}),\\
        &\mathbf{Q}_{<t,t} = \sum_{\mathbf{s}\in\mathcal{C}_t}\bphi_t(\mathbf{s})\bphi_{<t}^{\top}(\mathbf{s}),\\
        &\mathbf{Q}_{t,t} = \sum_{\mathbf{s}\in\mathcal{C}_t}\bphi_t(\mathbf{s})\bphi_t^{\top}(\mathbf{s}).
\end{align} 
The trace $\mathsf{Tr}\left(\mathbf{R}_t^{-1}\mathbf{J}_t\right)$ represents the prediction error reduction using the accumulated observations during FlyCom$^2$. Its maximization is directly determined by the optimization of $\bar{\mathbf{W}}_t$, while $\bar{\mathbf{W}}_t$ is coupled with the design of AirComp due to the power constraint:
\begin{align}
    P\geq \frac{\nu_{k,t}^2}{(\mathbf{b}_t^H\mathbf{h}_{k,t})^2}= \frac{\mathsf{Tr}\left(\mathbf{W}_{k,t}^{\top}\bPhi_{k,t}\mathbf{W}_{k,t}\right)}{N(\mathbf{b}_t^H\mathbf{h}_{k,t})^2},
\end{align}
where $\bPhi_{k,t} = \mathsf{E}[\mathbf{g}_{k,t}\mathbf{g}_{k,t}^{\top}]$. Hence, $\bar{\mathbf{W}}_t$ and $\mathbf{b}_t$ should be jointly designed to minimize the prediction error. Given $\mathbf{R}_t$ in~\eqref{eq:overall_cov} and $\mathbf{J}_t$ in~\eqref{eq:overall_pre}, the problem can be formulated as
\begin{equation}
    \begin{aligned}\label{prob:general_design}
    	\mathop{\max}_{\bar{\mathbf{W}}_t, \mathbf{b}_t}&\ \ \mathsf{Tr}\left(\mathbf{R}_t^{-1}\mathbf{J}_t\right)\\
     \mathrm{s.t.}
        &\ \ \bar{\mathbf{W}}_t = [\mathbf{W}_{1,t}^{\top},\ \mathbf{W}_{2,t}^{\top},\ \cdots,\mathbf{W}_{K,t}^{\top}]^{\top},\\
        &\ \ \mathsf{Tr}\left(\mathbf{W}_{k,t}^{\top}\bPhi_{k,t}\mathbf{W}_{k,t}\right)\leq NP(\mathbf{b}_t^H\mathbf{h}_{k,t})^2,\ \forall k.
    \end{aligned}
\end{equation}

\section{Communication-Computing Integrated Design for Distributed PtCloud Fusion}\label{section:design}
The coupling between communication and computing and the progressive property in FlyCom$^2$ makes it difficult to address the problem~\eqref{prob:general_design} directly. To this end, we first study two simplified cases induced from the FlyCom$^2$, namely distributed PtCloud fusion using one-shot noiseless aggregation and one-shot AirComp-based aggregation. Compared to the original one in~\eqref{prob:general_design}, these cases reveal a simplified form of problem formulation but involve some common tradeoffs between communication and computing, from which the obtained solutions and insights are exploited to facilitate the problem-solving for FlyCom$^2$.

\subsection{Case 1 -- One-Shot Noiseless PtCloud Fusion}
We first consider that the local observations $\{\mathbf{g}_{k,t}\}$ are generated from all local PtCloud points at each sensor and then are reliably aggregated from the sensors to the server. This is the ideal case for the distributed PtCloud fusion. In this case, the global fusion can be still performed via~\eqref{eq:global_predictor} with $\mathbf{R}_t = \bar{\mathbf{W}}^{\top}\bar{\bPhi}\bar{\mathbf{W}}$ and $\mathbf{J}_t=\bar{\mathbf{W}}^{\top}\mathbf{Q}\bar{\mathbf{W}}$, reducing the problem of local observation matrix desing to
\begin{equation}
    \begin{aligned}\label{prob:special_design}
    	\mathop{\max}_{\bar{\mathbf{W}}}&\ \ \mathsf{Tr}\left(\frac{\bar{\mathbf{W}}^{\top}\mathbf{Q}\bar{\mathbf{W}}}{\bar{\mathbf{W}}^{\top}\bar{\bPhi}\bar{\mathbf{W}}}\right).
    \end{aligned}
\end{equation}
The problem~\eqref{prob:special_design} reveals a form of \emph{ratio trace maximization} whose optimal solution, $\bar{\mathbf{W}}^{\star}$, can be obtained analytically by solving the generalized eigenvalue decomposition~\cite{statistic_learning_book}:
\begin{equation}\label{eq:GED}
    \mathbf{Q}\mathbf{w}_i = \lambda_i\bPhi\mathbf{w}_i,
\end{equation}
where $\lambda_i$ denotes the $i$-th largest generalized eigenvalue with $\mathbf{w}_i$ being its corresponding eigenvector. Importantly, we notice a strong connection between the observation matrix design and the well-known LDA problem that constructs a set of linear combinations of existing data points as low-dimensional observations to separate the classes of objects. The objective of LDA is to optimize the combination weights to simultaneously maximize the resulting inter-class variance/covariance, $\mathbf{\Sigma}_{\mathsf{inter}}$, and minimize intra-class variance/covariance, $\mathbf{\Sigma}_{\mathsf{intra}}$. The optimal result is obtained by solving: $\mathbf{\Sigma}_{\mathsf{inter}}\mathbf{w}=\lambda\mathbf{\Sigma}_{\mathsf{intra}}\mathbf{w}$, similar to~\eqref{eq:GED}.
\begin{Remark}\label{Remark:LDA}
    \emph{The local observation synthesis in the proposed FlyCom$^2$-based distributed PtCloud fusion is close to a process of LDA, which targets finding low-dimension observations, $\bar{\mathbf{f}}$, over data points scattered across multiple sensors. According to~\eqref{eq:GED}, the covariance of $\bar{\mathbf{f}}$ is given as $\mathsf{E}[\bar{\mathbf{f}}\bar{\mathbf{f}}^{\top}] =\bar{\mathbf{W}}^{\top}\bar{\bPhi}\bar{\mathbf{W}}$, which according to~\eqref{eq:GED}, will be optimized to be a diagonal matrix. This suggests that the observation matrix $\bar{\mathbf{W}}$ discards the potential correlation of data points between sensors to generate a set of independent observation samples for global prediction. The heterogeneity is addressed by manipulating the magnitude of each observation $\mathbf{W}_k$ in $\bar{\mathbf{W}}$ according to the spatial PtCloud correlation between sensors.  At the same time, $\mathbf{Q}$ reflects the spatial correlation between prediction areas and the obtained data points. Addressing~\eqref{prob:special_design} allows $\bar{\mathbf{W}}$ to align with the principal components of $\mathbf{Q}$ as much as possible, retaining the useful components in raw PtCloud points in on-sensor computation for the global prediction.}
\end{Remark}
Existing linear PtCloud compression methods (see, e.g.,~\cite{TIP_PCcompression}) are based on a process of \emph{principal component analysis} (PCA), designing $\bar{\mathbf{W}}_{k}$ for sensor $k$ as the principal eigenvectors of its correlation matrix $\mathbf{\Phi}_{k}$. According to~\eqref{prob:special_design}, they ignore the data heterogeneity between sensors and prediction importance, thereby being suboptimal for the current case even sharing the same form of local projection in~\eqref{eq:local_estimate}.

\subsection{Case 2 -- One-Shot Over-the-Air PtCloud Fusion}\label{subsec:oneshot}
Next, we consider the case of distributed PtCloud fusion with one-shot AirComp based aggregation of local observations. It is similar to the case of ideal distributed PtCloud fusion except for the additional absorption of channel distortions in fusion performance and the resulting AirComp receiver design. To be specific, based on~\eqref{prob:special_design} and~\eqref{prob:general_design}, the communication-computing integrated design in the current case can be particularized as
\begin{equation}
    \begin{aligned}\label{prob:oneshot_design}
    	\mathop{\max}_{\bar{\mathbf{W}}, \mathbf{b}}&\ \ \mathsf{Tr}\left(\frac{\bar{\mathbf{W}}^{\top}\mathbf{Q}\bar{\mathbf{W}}}{\bar{\mathbf{W}}^{\top}\bar{\bPhi}\bar{\mathbf{W}}+\frac{1}{2}\sigma^2\Vert\mathbf{b}\Vert^2\mathbf{I}_N}\right)\\
     \mathrm{s.t.}
        &\ \ \bar{\mathbf{W}} = [\mathbf{W}_1^{\top},\ \mathbf{W}_2^{\top},\ \cdots,\mathbf{W}_K^{\top}]^{\top}\\
        &\ \ \mathsf{Tr}\left(\mathbf{W}_k^{\top}\bPhi_k\mathbf{W}_k\right)\leq NP(\mathbf{b}^H\mathbf{h}_k)^2,\ \forall k.
    \end{aligned}
\end{equation}
To address the problem~\eqref{prob:oneshot_design}, we equivalently decompose it into two subsequent subproblems: 1) optimizing AirComp receiver $\mathbf{b}$ under arbitrary $\bar{\mathbf{W}}$ and 2) optimizing $\bar{\mathbf{W}}$ based on $\mathbf{b}^{\star}$, which are solved as follows.
\subsubsection{Optimization of AirComp receiver}
To begin with, we first give the following property and show the proof details in Appendix~\ref{Apdx:monotonicity}.
\begin{Lemma}\label{Lemma:monotonicity}
    \emph{Given two arbitrary positive-definite matrices denoted by $\mathbf{M}_1$ and $\mathbf{M}_2$, and a positive-semidefinite matrix $\mathbf{M}_0$, there is
    \begin{equation*}
        \mathsf{Tr}\left(\frac{\mathbf{M}_0}{\mathbf{M}_1}\right)\leq\mathsf{Tr}\left(\frac{\mathbf{M}_0}{\mathbf{M}_2}\right),\ \mathrm{if\ }\mathbf{M}_1\succeq\mathbf{M}_2.
    \end{equation*}}
\end{Lemma}
Lemma~\ref{Lemma:monotonicity} suggests that the optimization of $\mathbf{b}^{\star}$ is consistent with the conventional AirComp error minimization whose solution is given as~\cite{GXZhuAirComp2019}. Specifically, given arbitrary $\mathbf{b}$, the optimal $\mathbf{b}^{\star}$ is determined if and only if $\Vert\mathbf{b}\Vert^2 \leq \Vert\mathbf{b}^{\star}\Vert^2$, which ensures that for any $\bar{\mathbf{W}}$, there is
\begin{equation*}
    \bar{\mathbf{W}}^{\top}\bar{\bPhi}\bar{\mathbf{W}}+\frac{1}{2}\sigma^2\Vert\mathbf{b}^{\star}\Vert^2\mathbf{I}_N\succeq\bar{\mathbf{W}}^{\top}\bar{\bPhi}\bar{\mathbf{W}}+\frac{1}{2}\sigma^2\Vert\mathbf{b}\Vert^2\mathbf{I}_N,\ \forall \mathbf{b}.
\end{equation*}
The above property, together with Lemma~\ref{Lemma:monotonicity}, suggests that $\mathbf{b}^{\star}$ maximizes the trace value in~\eqref{prob:oneshot_design}. Therefore, according to~\cite{GXZhuAirComp2019}, the optimal solution for the AirComp receiver is given as
    \begin{equation}\label{eq:opt_AirComp}
        \mathbf{b}^{\star} = \sqrt{\max_k\ \frac{1}{NP}\mathsf{Tr}\left(\mathbf{\Psi}_k \right)}\cdot\mathbf{v},
    \end{equation}
    where $\mathbf{\Psi}_k \overset{\triangle}{=} \frac{1}{(\mathbf{v}^H\mathbf{h}_k)^2}\mathbf{W}_k^{\top}\bPhi_k\mathbf{W}_k$, and $\mathbf{v}$ is the first eigenvector of the channel matrix $\mathbf{H} = [\mathbf{h}_1,\mathbf{h}_2,\cdots,\mathbf{h}_K]$. Furthermore, with transmit SNR $\gamma = P/\sigma^2$, we have the power of effective AirComp noise expressible as
    \begin{equation}\label{eq:eff_noisepower}
        \frac{1}{2}\sigma^2\Vert\mathbf{b}^{\star}\Vert^2 = \frac{1}{2N\gamma}\max_k\ \mathsf{Tr}\left(\mathbf{\Psi}_k \right).
    \end{equation}

\subsubsection{Optimization of observation matrices}Using the result in~\eqref{eq:eff_noisepower}, the problem of designing $\bar{\mathbf{W}}$ is re-expressed from~\eqref{prob:oneshot_design} to
\begin{equation}\label{prob:design_W}
    \begin{aligned}
        \mathop{\max}_{\bar{\mathbf{W}}}&\ \ \mathsf{Tr}\left(\frac{\bar{\mathbf{W}}^{\top}\mathbf{Q}\bar{\mathbf{W}}}{\bar{\mathbf{W}}^{\top}\bar{\bPhi}\bar{\mathbf{W}}+\frac{1}{2N\gamma}\max_k\ \mathsf{Tr}\left(\mathbf{\Psi}_k \right)\mathbf{I}_N}\right)\\
     \mathrm{s.t.}
        &\ \ \bar{\mathbf{W}} = [\mathbf{W}_1^{\top},\ \mathbf{W}_2^{\top},\ \cdots,\mathbf{W}_K^{\top}]^{\top},\\
        &\ \ \mathbf{\Psi}_k = \frac{1}{(\mathbf{v}^H\mathbf{h}_k)^2}\mathbf{W}_k^{\top}\bPhi_k\mathbf{W}_k,\ \forall k.
    \end{aligned}
\end{equation}
It is difficult to optimally solve~\eqref{prob:design_W} due to the non-convex objective function. To overcome this problem, we resort to maximizing the lower or upper bound of the objective function, based on the following property.
\begin{Lemma}\label{Lemma:bound}
    \emph{Define $\mathbf{\Psi} = \frac{1}{N}\mathsf{diag}\left(
        \frac{\bPhi_1}{(\mathbf{v}^H\mathbf{h}_1)^2},\cdots,\frac{\bPhi_K}{(\mathbf{v}^H\mathbf{h}_K)^2}\right)$ and $k^* = \arg\max_k\ \mathsf{Tr}\left(\mathbf{\Psi}_k \right)$, there is
        \begin{equation}
           \frac{1}{\kappa}\bar{\mathbf{W}}^{\top}\mathbf{\Psi}\bar{\mathbf{W}}\succeq  \frac{1}{N}\max_k\ \mathsf{Tr}(\mathbf{\Psi}_k)\mathbf{I}_N\succeq \frac{1}{K}\bar{\mathbf{W}}^{\top}\mathbf{\Psi}\bar{\mathbf{W}},
        \end{equation}
        where $\kappa$ is an arbitrary positive value with $\kappa\leq \frac{\lambda_{\mathbf{\Psi}_{k^*},\min}}{\mathsf{Tr}\left(\mathbf{\Psi}_{k^*} \right)}$ and $\lambda_{\mathbf{\Psi}_{k^*},\min}$ being the smallest eigenvalue of $\mathbf{\Psi}_{k^*}$.}
\end{Lemma}
Detailed proofs are given in Appendix~\ref{Apdx:matrix_bound}. Using the Lemma~\ref{Lemma:monotonicity} and Lemma~\ref{Lemma:bound}, the objective function in problem~\eqref{prob:design_W} can be lower and upper bounded as
\begin{align}\label{eq:approximate}
    &\mathsf{Tr}\left(\frac{\bar{\mathbf{W}}^{\top}\mathbf{Q}\bar{\mathbf{W}}}{\bar{\mathbf{W}}^{\top}(\bar{\bPhi}+\frac{1}{2\gamma \kappa}\mathbf{\Psi})\bar{\mathbf{W}}}\right)\nonumber\\ &\leq\mathsf{Tr}\left(\frac{\bar{\mathbf{W}}^{\top}\mathbf{Q}\bar{\mathbf{W}}}{\bar{\mathbf{W}}^{\top}\bar{\bPhi}\bar{\mathbf{W}}+\frac{1}{2N\gamma}\max_k\ \mathsf{Tr}\left(\mathbf{\Psi}_k \right)\mathbf{I}_N}\right)\nonumber\\
    &\leq \mathsf{Tr}\left(\frac{\bar{\mathbf{W}}^{\top}\mathbf{Q}\bar{\mathbf{W}}}{\bar{\mathbf{W}}^{\top}(\bar{\bPhi}+\frac{1}{2\gamma K}\mathbf{\Psi})\bar{\mathbf{W}}}\right).
\end{align}
The left and right sides of~\eqref{eq:approximate} provide us an approximate of the objective function problem~\eqref{prob:design_W}, enabling the removal of constraints in the problem~\eqref{prob:design_W} and simplifying it as
\begin{equation}
    \begin{aligned}
        \mathop{\max}_{\bar{\mathbf{W}}}&\ \ \mathsf{Tr}\left(\frac{\bar{\mathbf{W}}^{\top}\mathbf{Q}\bar{\mathbf{W}}}{\bar{\mathbf{W}}^{\top}(\bar{\bPhi} + \alpha^{-1}\gamma^{-1}\mathbf{\Psi})\bar{\mathbf{W}}}\right).
    \end{aligned}
\end{equation}
where $\alpha$ is a constant that can be selected as $\alpha\in[2\kappa,2K]$. It is observed that the new problem is consistent with that in~\eqref{prob:special_design} except for replacing $\bar{\bPhi}$ by $\bar{\bPhi} + \alpha^{-1}\gamma^{-1}\mathbf{\Psi}$. Hence, its optimal solution can be obtained by solving the generalized eigenvalue decomposition:
\begin{equation}\label{eq:opt_solution_aircomp}
    \boxed{\mathbf{Q}\bar{\mathbf{W}} = (\bar{\bPhi}+ \alpha^{-1}\gamma^{-1}\mathbf{\Psi})\bar{\mathbf{W}}\mathbf{\Lambda}}.
\end{equation}
Furthermore, it is observed from~\eqref{eq:opt_solution_aircomp} that $\bar{\mathbf{W}}$ achieves a tradeoff between the heterogeneous data compression and AirComp distortion suppression. For the distributed PtCloud fusion with one-shot AirComp aggregation, there are two possible benchmarking designs of $\bar{\mathbf{W}}$: 1) it is designed through~\eqref{eq:GED} that focuses on heterogeneity-aware data compression, i.e. Remark~\ref{Remark:LDA}, and 2) to minimize the AirComp error~\eqref{eq:eff_noisepower}, $\mathbf{W}_{k}$ can be computed as the non-principal eigenvectors of $\mathbf{\Phi}_k$ so as to maximize the effective receive SNR. Both designs are suboptimal for the current case. Specifically, using the design in~\eqref{eq:GED} ignores the effects of channel diversity. Sensor $k$ can have largely uncorrelated PtCloud points but suffer a poor channel gain. In this case, solely increasing the value of $\mathbf{W}_{k}$ in  $\bar{\mathbf{W}}$ will enlarge the resulting AirComp channel distortions and thus degrade the global fusion performance. On the other hand, the AirComp-error-minimization design endows the aggregated observation with less information for the global prediction because of the non-principal component selection. Despite the minimal aggregation error, this leads to poor prediction performance.
\begin{Remark}[Balanced Tradeoff Between Data Heterogeneity and Channel Diversity]\label{Remark:tradeoff}
    \emph{To strike a balance between the PtCloud compression with data heterogeneity and AirComp noise suppression, the observation matrix is optimized through~\eqref{eq:opt_solution_aircomp}, where compared to~\eqref{eq:GED}, $\bar{\bPhi}$ is replaced by $\bar{\bPhi}+\alpha^{-1}\gamma^{-1}\mathbf{\Psi}$ with $\mathbf{\Psi}$ reflecting the diversity of the channel gains of sensors. If within a large transmit SNR $\gamma\rightarrow\infty$, the solution in~\eqref{eq:opt_solution_aircomp} reduces to that in~\eqref{eq:GED}.}
\end{Remark}

\subsection{Case 3 -- On-the-Fly PtCloud Fusion}
Eventually, we address the joint design problem for FlyCom$^2$-based distributed PtCloud fusion. It is worth mentioning that the optimal AirComp receiver remains unchanged under the consideration of FlyCom$^2$ due to the monotonicity in Lemma~\ref{Lemma:monotonicity} and the properties of block diagonal matrices. That is
\begin{equation}
    \mathbf{b}_t^{\star} =\sqrt{\max_k\ \frac{1}{NP}\mathsf{Tr}\left(\mathbf{\Psi}_{k,t} \right)}\cdot\mathbf{v}_t,
\end{equation}
where $\mathbf{\Psi}_{k,t} =\frac{1}{(\mathbf{v}_t^H\mathbf{h}_{k,t})^2}\mathbf{W}_{k,t}^{\top}\bPhi_{k,t}\mathbf{W}_{k,t}$, and $\mathbf{v}_t$ is the first eigenvector of the channel matrix $\mathbf{H}_t = [\mathbf{h}_{1,t},\mathbf{h}_{2,t},\cdots,\mathbf{h}_{K,t}]$.
Hence, we reuse the solution in~\eqref{eq:opt_AirComp} and only need to focus on the optimization of $\bar{\mathbf{W}}_t$ in the sequel. For simplicity of notation, we first define
\begin{align}\label{eq:define1}
\tilde{\mathbf{Q}}_t  &= \begin{bmatrix}
        \bar{\mathbf{W}}_{<t}^{\top}&\\
        &\mathbf{I}
    \end{bmatrix}\begin{bmatrix}
        \mathbf{Q}_{<t,<t}&\mathbf{Q}_{<t,t}\\
        \mathbf{Q}_{<t,t}^{\top}&\mathbf{Q}_{t,t}
    \end{bmatrix}
    \begin{bmatrix}
        \bar{\mathbf{W}}_{<t}&\\
        &\mathbf{I}
    \end{bmatrix}\nonumber\\
    &=\begin{bmatrix}
        \bar{\mathbf{W}}_{<t}^{\top}\mathbf{Q}_{<t,<t}\bar{\mathbf{W}}_{<t}&\bar{\mathbf{W}}_{<t}^{\top}\mathbf{Q}_{<t,t}\\
        \mathbf{Q}_{<t,t}^{\top}\bar{\mathbf{W}}_{<t}&\mathbf{Q}_{t,t}
    \end{bmatrix},
\end{align}
 and 
 \begin{align}\label{eq:define2}
     \tilde{\bPhi}_t &= \begin{bmatrix}
        \bar{\mathbf{W}}_{<t}^{\top}&\\
        &\mathbf{I}
    \end{bmatrix}\begin{bmatrix}
        \bar{\bPhi}_{<t,<t}&\bar{\bPhi}_{<t,t}\\
        \bar{\bPhi}_{<t,t}^{\top}&\bar{\bPhi}_{t,t}
    \end{bmatrix}\begin{bmatrix}
        \bar{\mathbf{W}}_{<t}&\\
        &\mathbf{I}
    \end{bmatrix}\nonumber\\
    &=\begin{bmatrix}
        \bar{\mathbf{W}}_{<t}^{\top}\bar{\bPhi}_{<t,<t}\bar{\mathbf{W}}_{<t}&\bar{\mathbf{W}}_{<t}^{\top}\bar{\bPhi}_{<t,t}\\
        \bar{\bPhi}_{<t,t}^{\top}\bar{\mathbf{W}}_{<t}&\bar{\bPhi}_{t,t}
    \end{bmatrix}.
 \end{align}
Using the definitions in~\eqref{eq:define1} and~\eqref{eq:define2}, the objective function of~\eqref{prob:general_design} is rewritten as 
\begin{equation*}
    \mathsf{Obj}(\bar{\mathbf{W}}_t)=\mathsf{Tr}\left(\frac{\mathsf{diag}(\mathbf{I},\bar{\mathbf{W}}_t^{\top})\tilde{\mathbf{Q}}_t\mathsf{diag}(\mathbf{I},\bar{\mathbf{W}}_t) }{\mathsf{diag}(\mathbf{I},\bar{\mathbf{W}}_t^{\top})\cdot\tilde{\bPhi}_t\cdot\mathsf{diag}(\mathbf{I},\bar{\mathbf{W}}_t)+\mathsf{diag}\left(
\mathbf{\Psi}_{<t},\ \max_k\ \frac{1}{2N\gamma}\mathsf{Tr}\left(\mathbf{\Psi}_{k,t} \right)\mathbf{I}_N\right)}\right)
\end{equation*}
with the problem~\eqref{prob:general_design} expressible as
\begin{equation}\label{prob:flycom_design}
    \begin{aligned}
    	\mathop{\max}_{\bar{\mathbf{W}}_t}&\ \ \mathsf{Obj}(\bar{\mathbf{W}}_t)\\
     \mathrm{s.t.}
        &\ \ \bar{\mathbf{W}}_t = [\mathbf{W}_{1,t}^{\top},\ \mathbf{W}_{2,t}^{\top},\ \cdots,\mathbf{W}_{K,t}^{\top}]^{\top},\\
        &\ \ \mathbf{\Psi}_{k,t} =\frac{1}{(\mathbf{v}_t^H\mathbf{h}_{k,t})^2}\mathbf{W}_{k,t}^{\top}\bPhi_{k,t}\mathbf{W}_{k,t},\ \forall k.
    \end{aligned}
\end{equation}
According to Lemma~\ref{Lemma:bound}, $\mathsf{Obj}(\bar{\mathbf{W}}_t)$ can be bounded as
\begin{align}\label{eq:up_lw_bound}
        &\mathsf{Tr}\left(\frac{\mathsf{diag}(\mathbf{I},\bar{\mathbf{W}}_t^{\top})\cdot\tilde{\mathbf{Q}}_t\cdot\mathsf{diag}(\mathbf{I},\bar{\mathbf{W}}_t)}{\mathsf{diag}(\mathbf{I},\bar{\mathbf{W}}_t^{\top})\cdot\tilde{\mathbf{P}}_t(\kappa)\cdot\mathsf{diag}(\mathbf{I},\bar{\mathbf{W}}_t)}\right) \nonumber\\
        &\leq \mathsf{Obj}(\bar{\mathbf{W}}_t)\nonumber\\
        &\leq \mathsf{Tr}\left(\frac{\mathsf{diag}(\mathbf{I},\bar{\mathbf{W}}_t^{\top})\cdot\tilde{\mathbf{Q}}_t\cdot\mathsf{diag}(\mathbf{I},\bar{\mathbf{W}}_t)}{\mathsf{diag}(\mathbf{I},\bar{\mathbf{W}}_t^{\top})\cdot\tilde{\mathbf{P}}_t(K)\cdot\mathsf{diag}(\mathbf{I},\bar{\mathbf{W}}_t)}\right),
\end{align}
where $\tilde{\mathbf{P}}_t(\alpha)  = \tilde{\bPhi}_t+\mathsf{diag}\left(
        \mathbf{\Psi}_{<t},\ \frac{1}{\alpha\gamma}\mathbf{\Psi}_t\right)$.
This allows for the tractable approximation of the problem~\eqref{prob:flycom_design}:
\begin{equation}\label{prob:flycom_design}
    \begin{aligned}
    	\mathop{\max}_{\bar{\mathbf{W}}_t}&\ \ \mathsf{Tr}\left(\frac{\mathsf{diag}(\mathbf{I},\bar{\mathbf{W}}_t^{\top})\cdot\tilde{\mathbf{Q}}_t\cdot\mathsf{diag}(\mathbf{I},\bar{\mathbf{W}}_t)}{\mathsf{diag}(\mathbf{I},\bar{\mathbf{W}}_t^{\top})\cdot\tilde{\mathbf{P}}_t(\alpha)\cdot\mathsf{diag}(\mathbf{I},\bar{\mathbf{W}}_t)}\right),
    \end{aligned}
\end{equation}
which can be solved optimally with a closed-form solution by using the following identity.
\begin{Lemma}\label{Lemma:identity}
    \emph{Given arbitrary Hermitian matrices $\mathbf{Q} = \begin{bmatrix}
        \mathbf{Q}_1&\mathbf{Q}_2\\
        \mathbf{Q}_2^{\top}&\mathbf{Q}_3
    \end{bmatrix}$ and $\mathbf{P} =\begin{bmatrix}
        \mathbf{P}_1&\mathbf{P}_2\\
        \mathbf{P}_2^{\top}&\mathbf{P}_3
    \end{bmatrix}$, there exists an identity of
    \begin{equation}\label{eq:identity}
      \begin{aligned}
        \mathsf{Tr}\left(\frac{\mathsf{diag}(\mathbf{I},\mathbf{W}^{\top})\cdot\mathbf{Q}\cdot\mathsf{diag}(\mathbf{I},\mathbf{W})}{\mathsf{diag}(\mathbf{I},\mathbf{W}^{\top})\cdot\mathbf{P}\cdot\mathsf{diag}(\mathbf{I},\mathbf{W})}\right)=\mathsf{Tr}\left(\frac{\mathbf{W}^{\top}(\mathbf{Q}_3-\mathbf{\Xi})\mathbf{W}}{\mathbf{W}^{\top}(\mathbf{P}_3-\mathbf{\Upsilon})\mathbf{W}}\right) + \mathsf{Tr}\left(\frac{\mathbf{Q}_1}{\mathbf{P}_1}\right),
      \end{aligned}
    \end{equation}
    where the matrices of $\mathbf{\Upsilon}$ and $\mathbf{\Xi}$ are given as $\mathbf{\Upsilon} \overset{\triangle}{=} \mathbf{P}_2^{\top}(\mathbf{P}_1)^{-1}\mathbf{P}_2$ and $\mathbf{\Xi} \overset{\triangle}{=} \mathbf{Q}_2^{\top}(\mathbf{P}_1)^{-1}\mathbf{P}_2 + \mathbf{P}_2^{\top}(\mathbf{P}_1)^{-1}\mathbf{Q}_2-\mathbf{P}_2^{\top}(\mathbf{P}_1)^{-1}\mathbf{Q}_1(\mathbf{P}_1)^{-1}\mathbf{P}_2 $.}
\end{Lemma}

\begin{figure*}[!t]
 \normalsize 
    \begin{align}
    \mathbf{\Upsilon}_{<t,t} &= \bar{\bPhi}_{<t,t}^{\top}\bar{\mathbf{W}}_{<t}[\bar{\mathbf{W}}_{<t}^{\top}\bar{\bPhi}_{<t,<t}\bar{\mathbf{W}}_{<t} + \mathbf{\Psi}_{<t}]^{-1}\bar{\mathbf{W}}_{<t}^{\top}\bar{\bPhi}_{<t,t} \overset{\mathrm{Eq.}~\eqref{eq:overall_cov}}{=}\bar{\bPhi}_{<t,t}^{\top}\bar{\mathbf{W}}_{<t}\mathbf{R}_{t-1}^{-1}\bar{\mathbf{W}}_{<t}^{\top}\bar{\bPhi}_{<t,t},\label{eq:upsilon}\\
    \mathbf{\Xi}_{<t,t} &=
        \mathbf{Q}_{<t,t}^{\top}\bar{\mathbf{W}}_{<t}\mathbf{R}_{t-1}^{-1}\bar{\mathbf{W}}_{<t}^{\top}\bar{\bPhi}_{<t,t}+\bar{\bPhi}_{<t,t}^{\top}\bar{\mathbf{W}}_{<t}\mathbf{R}_{t-1}^{-1}\bar{\mathbf{W}}_{<t}^{\top}\mathbf{Q}_{<t,t}\nonumber\\
        &\quad-\bar{\bPhi}_{<t,t}^{\top}\bar{\mathbf{W}}_{<t}\mathbf{R}_{t-1}^{-1}\mathbf{J}_{t-1}\mathbf{R}_{t-1}^{-1}\bar{\mathbf{W}}_{<t}^{\top}\bar{\bPhi}_{<t,t}.\label{eq:xi}
    \end{align}
     \hrulefill
\end{figure*}
The detailed proof is given in Appendix~\ref{Apdx:identity}. Substituing Lemma~\ref{Lemma:identity} in the problem~\eqref{prob:flycom_design}, one can notice that the second term in the right side of the identity~\eqref{eq:identity}, $\mathsf{Tr}\left(\frac{\mathbf{Q}_1}{\mathbf{P}_1}\right)$, is independent of $\bar{\mathbf{W}}_t$. Actually, $\mathsf{Tr}\left(\frac{\mathbf{Q}_1}{\mathbf{P}_1}\right)$ represents the cumulative error reduction to the previous slot, which is shown in the later performance analysis. Therefore, the minimization of the estimation error can be simplified to maximize the first term in the identity~\eqref{eq:identity}, which is expressible as
\begin{equation}\label{prob:final_case}
    \begin{aligned}
    	\mathop{\max}_{\bar{\mathbf{W}}_t}&\ \ \mathsf{Tr}\left(\frac{\bar{\mathbf{W}}_t^{\top}(\mathbf{Q}_{t,t}-\mathbf{\Xi}_{<t,t})\bar{\mathbf{W}}_t}{\bar{\mathbf{W}}_t^{\top}(\bar{\bPhi}_{t,t} + \alpha^{-1}\gamma^{-1}\mathbf{\Psi}_t-\mathbf{\Upsilon}_{<t,t})\bar{\mathbf{W}}_t}\right),
    \end{aligned}
\end{equation}
where $\mathbf{\Upsilon}_{<t,t}$ and $\mathbf{\Xi}_{<t,t}$ are given in~\eqref{eq:upsilon} and~\eqref{eq:xi}, respectively. It is observed from~\eqref{prob:final_case} that \eqref{prob:final_case} still shares the same form with~\eqref{eq:GED} except for a changed decomposable matrix. Therefore, the optimal observation matrix for FlyCom$^2$ is obtained by solving the corresponding generalized eigenvalue decomposition:
\begin{equation}\label{eq:opt_solution_flycom}
    \boxed{(\mathbf{Q}_{t,t}\!-\!\mathbf{\Xi}_{<t,t})\bar{\mathbf{W}}_t = \left(\bar{\bPhi}_{t,t} \!+\!\frac{1}{\alpha\gamma}\mathbf{\Psi}_t\!-\!\mathbf{\Upsilon}_{<t,t}\right)\bar{\mathbf{W}}_t\mathbf{\Lambda}_t}.
\end{equation}
It should be emphasized that the left and right decomposable matrices in~\eqref{eq:opt_solution_flycom}, $\mathbf{Q}_{t,t}\!-\!\mathbf{\Xi}_{<t,t}$ and $\bar{\bPhi}_{t,t} +\frac{1}{\alpha\gamma}\mathbf{\Psi}_t-\mathbf{\Upsilon}_{<t,t}$, will not necessarily be positive-definite matrices. Therefore, the generalized eigenvectors, as well as their corresponding eigenvalues, could be composed of complex values. Only the real eigenvectors satisfy the real constraints of $\bar{\mathbf{W}}_t$ and will be used as columns in $\bar{\mathbf{W}}_t$. Furthermore, the achieved objective function with the optimal design is expressible as
\begin{align}\label{eq:temporal_gain}
    &\mathsf{Tr}\left(\frac{\bar{\mathbf{W}}_t^{\top}(\mathbf{Q}_{t,t}-\mathbf{\Xi}_{<t,t})\bar{\mathbf{W}}_t}{\bar{\mathbf{W}}_t^{\top}(\bar{\bPhi}_{t,t} + \alpha^{-1}\gamma^{-1}\mathbf{\Psi}_t-\mathbf{\Upsilon}_{<t,t})\bar{\mathbf{W}}_t}\right) \nonumber\\
    &= \mathsf{Tr}\left(\frac{\bar{\mathbf{W}}_t^{\top}(\bar{\bPhi}_{t,t} + \alpha^{-1}\gamma^{-1}\mathbf{\Psi}_t-\mathbf{\Upsilon}_{<t,t})\bar{\mathbf{W}}_t\mathbf{\Lambda}_t}{\bar{\mathbf{W}}_t^{\top}(\bar{\bPhi}_{t,t} + \alpha^{-1}\gamma^{-1}\mathbf{\Psi}_t-\mathbf{\Upsilon}_{<t,t})\bar{\mathbf{W}}_t}\right) \nonumber\\
    & = \mathsf{Tr}\left(\mathbf{\Lambda}_t\right),
\end{align}
where according to Lemma~\ref{Lemma:identity} and~\eqref{eq:temporal_error}, $\mathsf{Tr}\left(\mathbf{\Lambda}_t\right)$ represents the prediction error reduction in the current slot. Hence, to maximize the gain of prediction error reduction, the generalized eigenvectors in~\eqref{eq:opt_solution_flycom} corresponding to positive eigenvalues are selected and included in $\bar{\mathbf{W}}_t$.

\begin{table*}[h]
\caption{Designing observation matrices within different cases via generalized eigenvalue decomposition: $\mathbf{Q}\mathbf{W} = \bPhi \mathbf{W}\mathbf{\Lambda}$}
\centering
\footnotesize
{\renewcommand{\arraystretch}{1.5}
\begin{tabular}{|c|ccc|}
\hline
& \multicolumn{1}{c|}{$\mathbf{Q}=$} & \multicolumn{1}{c|}{$\mathbf{\Phi}=$} &\multicolumn{1}{c|}{Interpretation}
\\ \hline
Ideal distributed PtCloud fusion     & \multicolumn{1}{c|}{$\int_{\mathcal{C}}\bphi(\mathbf{s})\bphi^{\top}(\mathbf{s})\mathrm{d}\mathbf{s}$} & \multicolumn{1}{c|}{$\mathsf{E}\left[\bar{\mathbf{g}}\bar{\mathbf{g}}^{\top}\right]$} &\multicolumn{1}{c|}{Remark~\ref{Remark:LDA}}  \\ \hline
AirComp-based distributed PtCloud fusion & \multicolumn{1}{c|}{$\int_{\mathcal{C}}\bphi(\mathbf{s})\bphi^{\top}(\mathbf{s})\mathrm{d}\mathbf{s}$} & \multicolumn{1}{c|}{$\mathsf{E}\left[\bar{\mathbf{g}}\bar{\mathbf{g}}^{\top}\right] + \gamma^{-1}\alpha^{-1}\mathbf{\Psi}$} &\multicolumn{1}{c|}{Remark~\ref{Remark:tradeoff}} \\ \hline
FlyCom$^2$-based distributed PtCloud fusion           & \multicolumn{1}{c|}{$\int_{\mathcal{C}}\bphi_t(\mathbf{s})\bphi_t^{\top}(\mathbf{s})\mathrm{d}\mathbf{s}-\mathbf{\Xi}_{<t,t}$} & \multicolumn{1}{c|}{$
    \mathsf{E}\left[\bar{\mathbf{g}}_t\bar{\mathbf{g}}_t^{\top}\right] + \gamma^{-1}\alpha^{-1}\mathbf{\Psi}_t-\mathbf{\Upsilon}_{<t,t}$} &\multicolumn{1}{c|}{Remark~\ref{Remark:uncorrelated}}  \\ \hline
\end{tabular}
\label{tbl:conclusion}}
\end{table*}
\begin{Remark}[Temporal Correlation in FlyCom$^2$]\label{Remark:uncorrelated}
    \emph{The correlation between current and previous observations leads to the design of FlyCom$^2$-based distributed PtCloud fusion distinct from its one-shot counterpart. Specifically, the optimal design for FlyCom$^2$-based distributed PtCloud fusion needs to further discard the components in current samples that reveal a strong correlation with previously received observations. This is reflected by the subtraction of $\mathbf{\Upsilon}_{<t,t}$ from $\bar{\bPhi}_{t,t}$. At the same time, it also takes into account the prediction performance using previous results to focus on these areas still with limited prediction performance, as reflected by the subtraction of $\mathbf{\Xi}_{<t,t}$ from $\mathbf{Q}_{t,t}$.}
\end{Remark}
Obviously, if the current observations are independent of previous samples, namely $\tilde{\bPhi}_t$ is a block diagonal matrix, there are $\bar{\bPhi}_{<t,t}=\mathbf{0}$, $\mathbf{\Upsilon}_{<t,t} = \mathbf{0}$, and $\mathbf{\Xi}_{<t,t} = \mathbf{0}$ in~\eqref{eq:opt_solution_flycom}. This will lead to the observation matrix optimized for FlyCom$^2$ reducing into that for one-shot aggregation. The designs of the above three cases are concluded in Table~\ref{tbl:conclusion} for quick comparison.

\section{Performance Analysis and Termination Rule for FlyCom$^2$}\label{section:stopping}
The prediction performance will be shown to continuously improve as the FlyCom$^2$-based distributed PtCloud fusion progresses and then eventually converge into a fixed value, based on which we discuss the termination of the process of streaming operations. 

\begin{figure}[t]
     \centering
    \subfigure[Value of the real-valued eigenvalues]{\label{subfig:distribution}\includegraphics[width=0.45\textwidth]{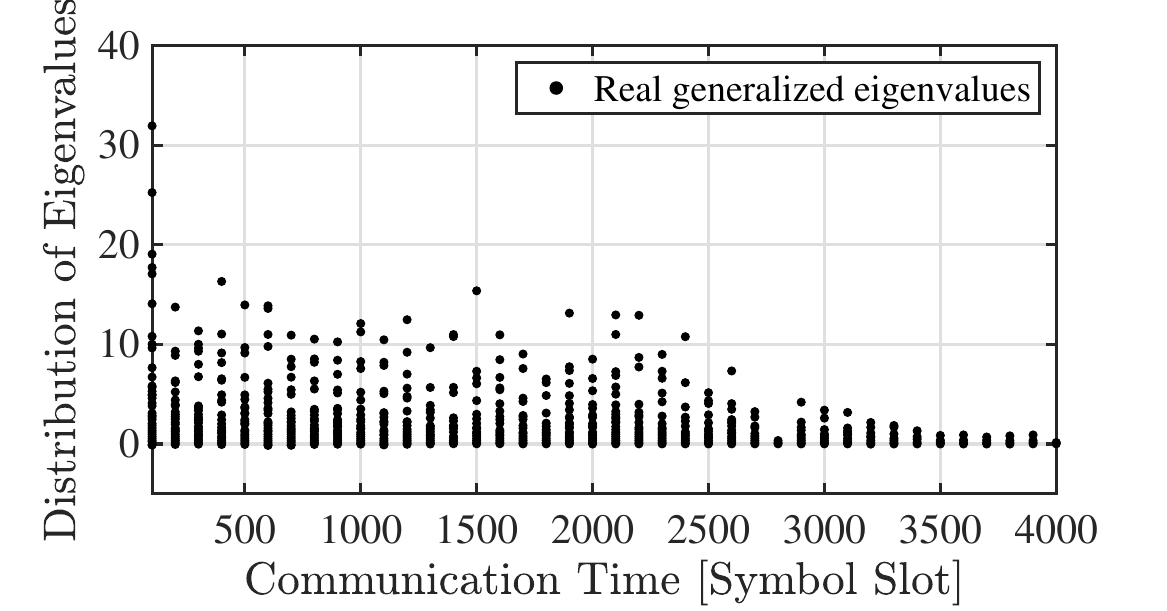}}
    \subfigure[Number of the real-valued positive eigenvalues]{\label{subfig:population}\includegraphics[width=0.45\textwidth]{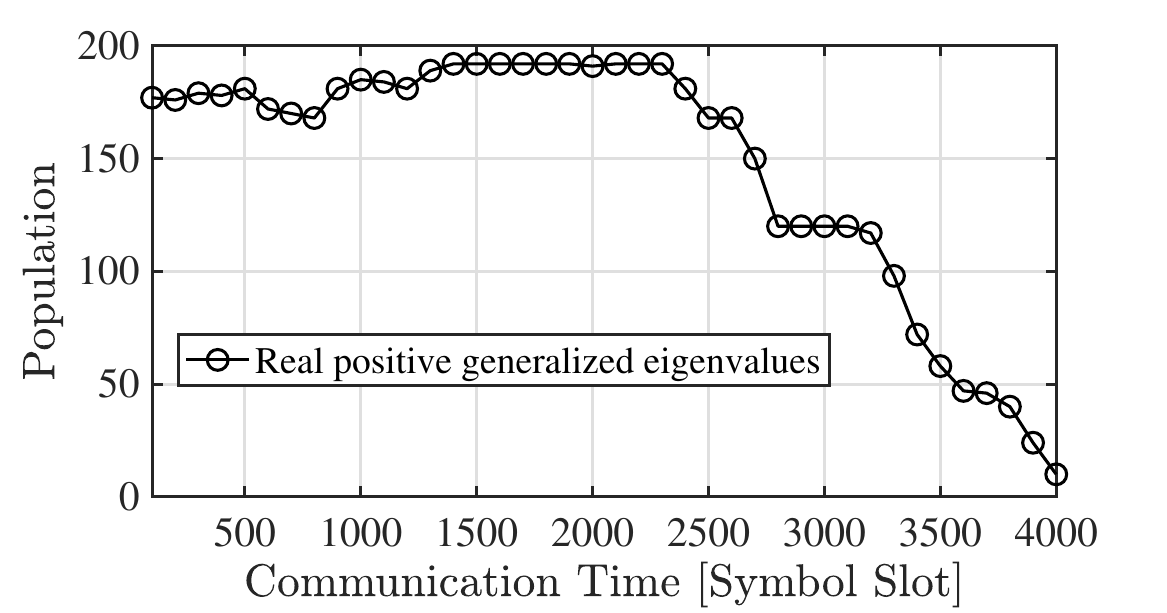}}
    \caption{Generalized eigenvalues obtained through GED in~\eqref{eq:opt_solution_flycom} for the progressive FlyCom$^2$ operations. Data samples are selected from depth-$5$ octree nodes. Each sensor will process $N=24$ bins. Other experimental settings are the same as those discussed in Section~\ref{section:experiments}.}
    \label{fig:eig_distribution}
\end{figure}
To this end, we investigate the prediction error caused by the designs~\eqref{eq:opt_solution_flycom} as follows. First, let rewrite the $\mathsf{Tr}\left(\frac{\mathbf{Q}_1}{\mathbf{P}_1}\right)$ in Lemma~\ref{Lemma:identity} according to the definitions of $\mathbf{Q}_1$ and $\mathbf{P}_1$ as
\begin{align}\label{eq:accumulated_reduction}
     \mathsf{Tr}\left(\frac{\mathbf{Q}_1}{\mathbf{P}_1}\right)
     &= \mathsf{Tr}\left(\frac{\bar{\mathbf{W}}_{<t}^{\top}\mathbf{Q}_{<t,<t}\bar{\mathbf{W}}_{<t}}{\bar{\mathbf{W}}_{<t}^{\top}(\bar{\bPhi}_{<t,<t} + \alpha^{-1}\gamma^{-1}\mathbf{\Psi}_{<t})\bar{\mathbf{W}}_{<t}}\right)\nonumber\\
    & \overset{(a)}{=} \mathsf{Tr}\left(\frac{\bar{\mathbf{W}}_{<t-1}^{\top}\mathbf{Q}_{<t-1,<t-1}\bar{\mathbf{W}}_{<t-1}}{\bar{\mathbf{W}}_{<t-1}^{\top}(\bar{\bPhi}_{<t-1,<t-1} + \alpha^{-1}\gamma^{-1}\mathbf{\Psi}_{<t-1})\bar{\mathbf{W}}_{<t-1}}\right)+ \mathsf{Tr}\left(\mathbf{\Lambda}_{t-1}\right)\nonumber\\
    &\overset{(b)}{=}\sum_{\ell\leq t-1}\mathsf{Tr}\left(\mathbf{\Lambda}_{\ell}\right),
\end{align}
where step (a) follows from~\eqref{eq:temporal_gain} and the identity in Lemma~\ref{Lemma:identity} and step (b) repeats the recursion process to obtain the final result. Therefore, define $\mathbf{\Lambda}_t^{\mathsf{up}}$ and $\mathbf{\Lambda}_t^{\mathsf{lw}}$ representing the eigenvalues in~\eqref{eq:temporal_gain} with $\alpha = K$ and $\alpha=\kappa$ respectively. Using the lower and upper bounds in~\eqref{eq:up_lw_bound} and the accumulation reduction~\eqref{eq:accumulated_reduction}, one can get the error in~\eqref{eq:temporal_error} bounded as
\begin{equation}\label{eq:error_bound}
    \beta - \frac{1}{|\mathcal{C}|}\sum_{\ell\leq t}\mathsf{Tr}\left(\mathbf{\Lambda}_t^{\mathsf{up}}\right)\leq\mathsf{error}_t\leq\beta - \frac{1}{|\mathcal{C}|}\sum_{\ell\leq t}\mathsf{Tr}\left(\mathbf{\Lambda}_t^{\mathsf{lw}}\right).
\end{equation}
The results in \eqref{eq:error_bound} suggest that the prediction error is both upper and lower bounded by two monotonic curves that decrease with respect to $t$, and in a single slot, the error reduction is given as the summation of all positive eigenvalues obtained in~\eqref{eq:opt_solution_flycom}. Furthermore, the prediction error reduction will converge as the FlyCom$^2$ progresses. According to~\eqref{eq:xi}, there is $\mathbf{Q}_{t,t}-\mathbf{\Xi}_{<t,t}\preceq\mathbf{0}$ with  $t\rightarrow\infty$, suggesting that the underpredicted areas using the accumulated observations are close to zero. This leads to the dimensionality of the current observation matrix close to zero, i.e., $\mathsf{Dim}\left(\bar{\mathbf{W}}_t\right) = 0,\ t\rightarrow \infty$ and
\begin{equation}
    \mathsf{Tr}\left(\mathbf{\Lambda}_t^{\mathsf{lw}}\right)=\mathsf{Tr}\left(\mathbf{\Lambda}_t^{\mathsf{up}}\right) =0,\ t\rightarrow \infty.
\end{equation}
The above degradation property allows for the convergence of the trace summation in~\eqref{eq:error_bound}. 
This is evaluated by the numerical simulations in Fig.~\ref{fig:eig_distribution} that reveals the distribution of the generalized eigenvalues, $\mathbf{\Lambda}_t$, generated through~\eqref{eq:opt_solution_flycom} within the progressive FlyCom$^2$ operations. It can be observed from Fig.~\ref{subfig:distribution} that as FlyCom$^2$ progresses, the eigenvalues are gradually concentrated to zero, of which the number of the positive eigenvalues, as shown in Fig.~\ref{subfig:population}, also reduces to zero.

The above discussion offers a natural termination rule as follows. 
\begin{Proposition}[Termination Rule]
\emph{The progressive operations of the FlyCom$^2$-based distributed PtCloud fusion proposed in Section~\ref{section:protocol} can terminate when the dimensionality of the observation matrix designed in~\eqref{eq:opt_solution_flycom} converges to zero.}
\end{Proposition}

\section{Experiments}\label{section:experiments}
\subsection{Experimental Setting}
Consider the ISEA system as shown in Fig.~\ref{fig:system_model}. The receive array size is $M = 8$. Assuming frequency non-selective Rayleigh fading, the multi-access channel vectors are composed of i.i.d. complex Gaussian $\mathcal{CN}(0,1)$ elements. Each communication slot is considered to spread over at least $I=100$ symbol intervals (smaller than a channel coherence block~\cite{Bjornson2016Tenmyths}) to support the transmission of uncoded symbols in AirComp. The transmit SNR is $\gamma =\{5,15\}$ dB.

\subsubsection{Experimental datasets}
\begin{figure}
     \centering
    \subfigure[PtCloud of \emph{Andrew}]{\includegraphics[width=0.34\textwidth]{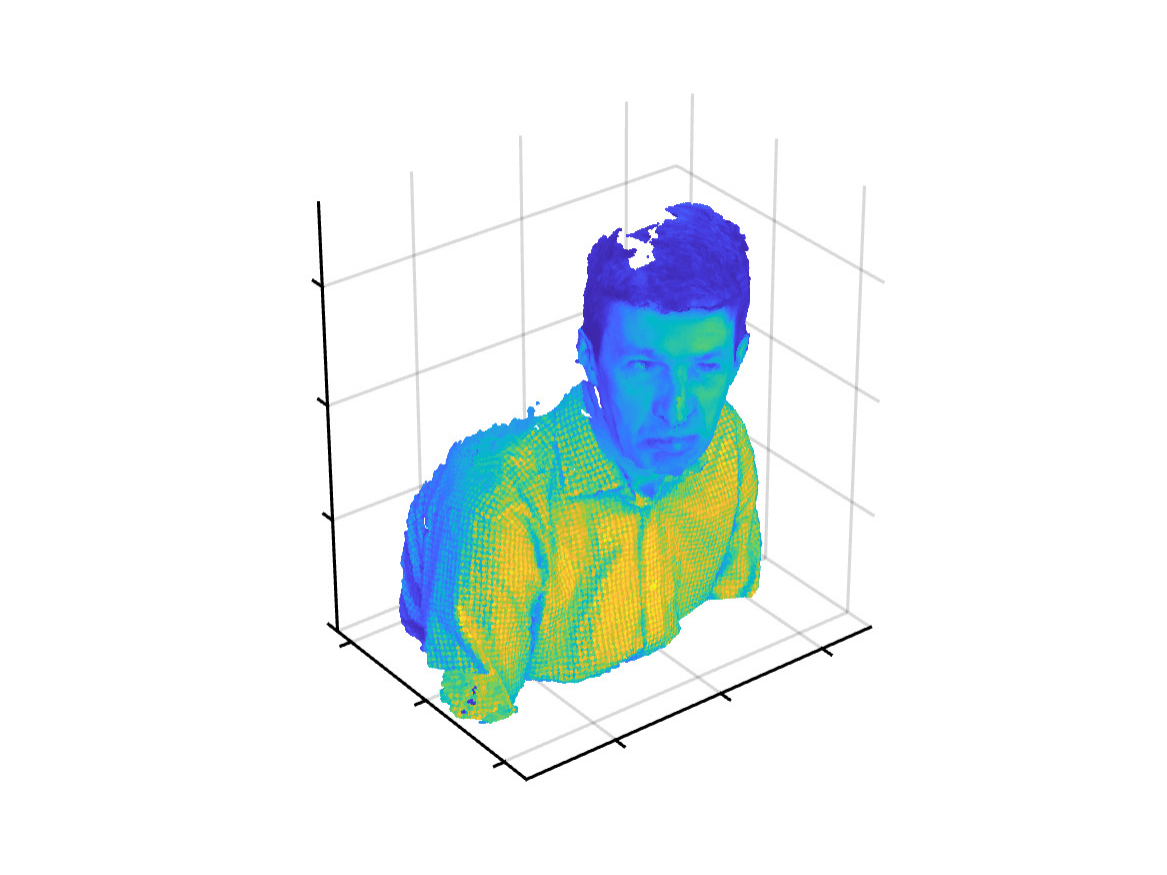}
    \label{subfig:andrew}
    }
    \subfigure[Distributed PtClouds]{\includegraphics[width=0.48\textwidth]{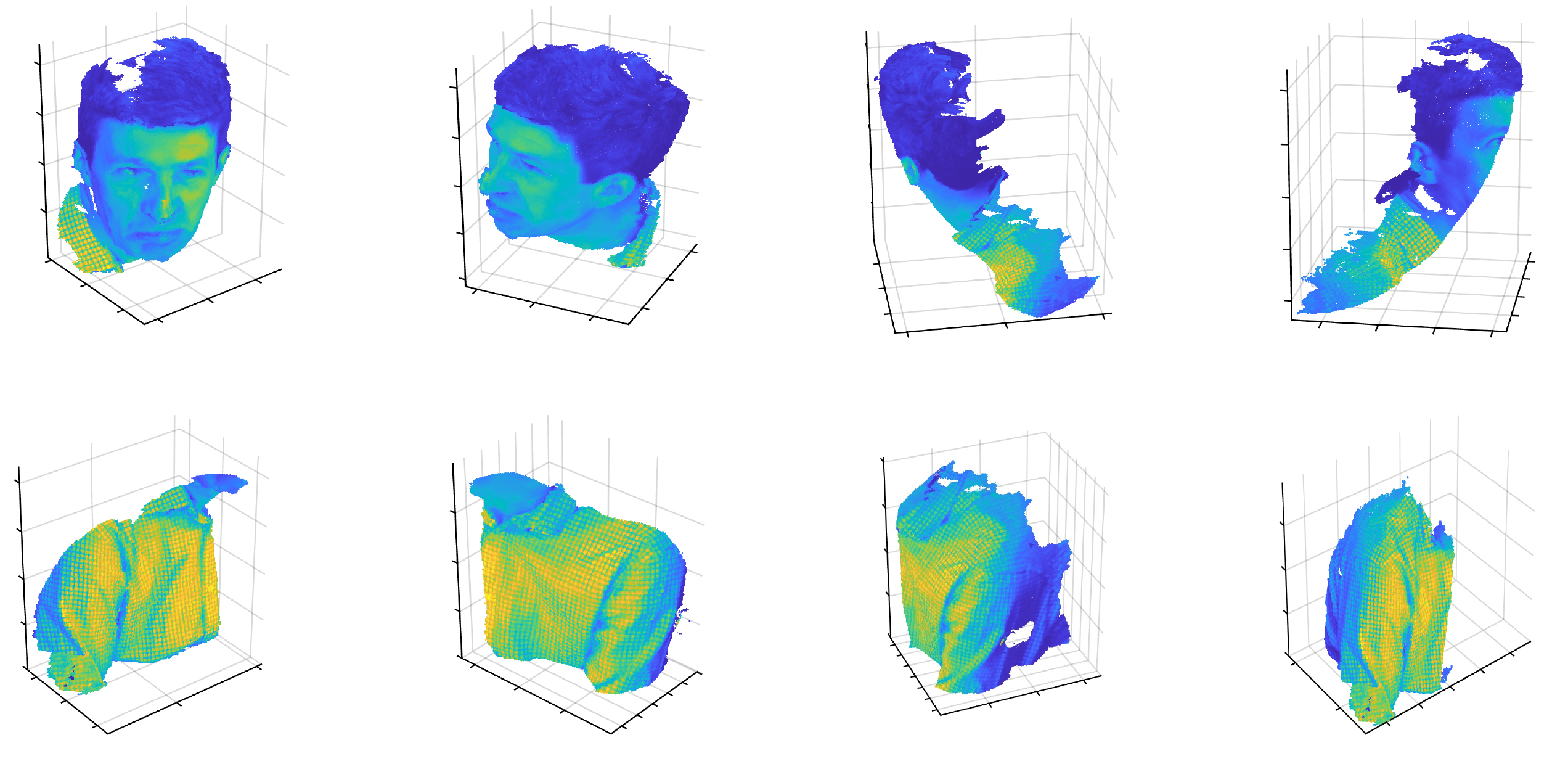}}
    \caption{Illustration of Microsoft Voxelized Upper Bodies and their distributed PtClouds.}
    \label{fig:data_show}
\end{figure}

We evaluate the proposed FlyCom$^2$-based distributed PtCloud fusion using the PtCloud datasets of Microsoft Voxelized Upper Bodies, known as \emph{Andrew}, \emph{David}, \emph{Phil}, \emph{Ricardo}, and \emph{Sarah}, that are captured by a real-time telepresence system with RGB-D sensors~\cite{MVUB_data}. Herein, each PtCloud in the original dataset is associated with RGB colors. For simplicity, we only focus on the luminance components and convert the $3$-dimensional color attributes to grayscales using the standardized luminance formula: $Grayscale = 0.299R + 0.587G + 0.114B$. To apply AirComp-based transmission into distributed PtCloud fusion, the local PtCloud attributes are further normalized to be zero-mean with unit variance. Following~\cite{TIP_PCcompression}, a squared exponential kernel is leveraged in the Gaussian process modeling that describes the inherent spatial correlations among PtCloud attributes. The required hyperparameters used in Gaussian process are learned as $\beta = 0.03$, $\sigma_{\sf e}^2 = 0$, and $\theta = 0.95$~\cite{TIP_PCcompression}. Finally, the original PtCloud dataset is not a multi-view version. Mimicking the multi-view dataset generation commonly used in the literature (see, e.g.,~\cite{MVCNN}), we generate the distributed PtCloud datasets from the original datasets by implementing $K=8$ virtual cameras with limited fields of view and detection range to capture partially the entire PtCloud from different angles. The distributed PtCloud datasets are illustrated in Fig.~\ref{fig:data_show}. 
\subsubsection{Benchmarking schemes}
We compare the performance of the proposed FlyCom$^2$ framework with three benchmarking schemes that support progressive PtCloud fusion similar to the proposed FlyCom$^2$. They can be treated as variants of FlyCom$^2$, comprising local observation synthesis, AirComp-based aggregation, and global regression, except for a distinct design of local observation matrices. The details are given in the sequel.
\begin{itemize}
    \item \textbf{PCA-FlyCom$^2$:} PCA-FlyCom$^2$ is to extend an representative one-shot PtCloud compression approach~\cite{TIP_PCcompression} to the FlyCom$^2$ framework. It achieves a PtCloud compression efficiency similar to other linear PtCloud compression methods in the literature, e.g., HoloCast~\cite{HoloCast_1} and HoloCast+~\cite{HoloCast_2}.
    In PCA-FlyCom$^2$, local observations are still generated at sensors through linear projection, where the observation matrices are individually computed as the principal components of the spatial covariance matrix of the on-sensor PtClouds~\cite{TIP_PCcompression}. Then, these observations are aggregated at the server via AirComp~\eqref{eq:AirComp_agg} and are then used to perform a global regression~\eqref{eq:global_predictor}.
     
    \item \textbf{AirComp-Error-Minimized (A.E.M.) FlyCom$^2$:} In contrast to PCA-FlyCom$^2$, the current scheme designs the local observation matrix as the non-principal component of the PtCloud covariance matrix, which can maximize the effective receive SNR in AirComp according to~\eqref{eq:eff_noisepower} and thus optimize the AirComp aggregation accuracy using the centroid receiver~\cite{Aggregationgain_JSAC,GXZhuAirComp2019}. 
    \item \textbf{Noncausal FlyCom$^2$:} Noncausal FlyCom$^2$ is a variant of the proposed FlyCom$^2$ algorithm while treating the involved progressive communication and computing as a set of independent operations. That is, in each current slot, the observation matrix and AirComp receiver design are jointly optimized to balance the tradeoff between data heterogeneity and AirComp distortions in a one-shot manner as in Section~\ref{subsec:oneshot}. Given the currently received observation and previous accumulations, a global regression is executed based on~\eqref{eq:global_predictor} for PtCloud fusion. 
\end{itemize}
Finally, we also present the PtCloud rendering results within distributed PtCloud fusion using the proposed FlyCom$^2$ method and compare them with the rendering using Oracle fusion that aggregates all the raw PtCloud points from sensors to form a union at the server~\cite{Cooper}.

\subsection{Error Evaluation of FlyCom$^2$-Enabled Distributed PtCloud Fusion}
\begin{figure}[t]
    \centering
    \subfigure[]{\includegraphics[width=0.45\textwidth]{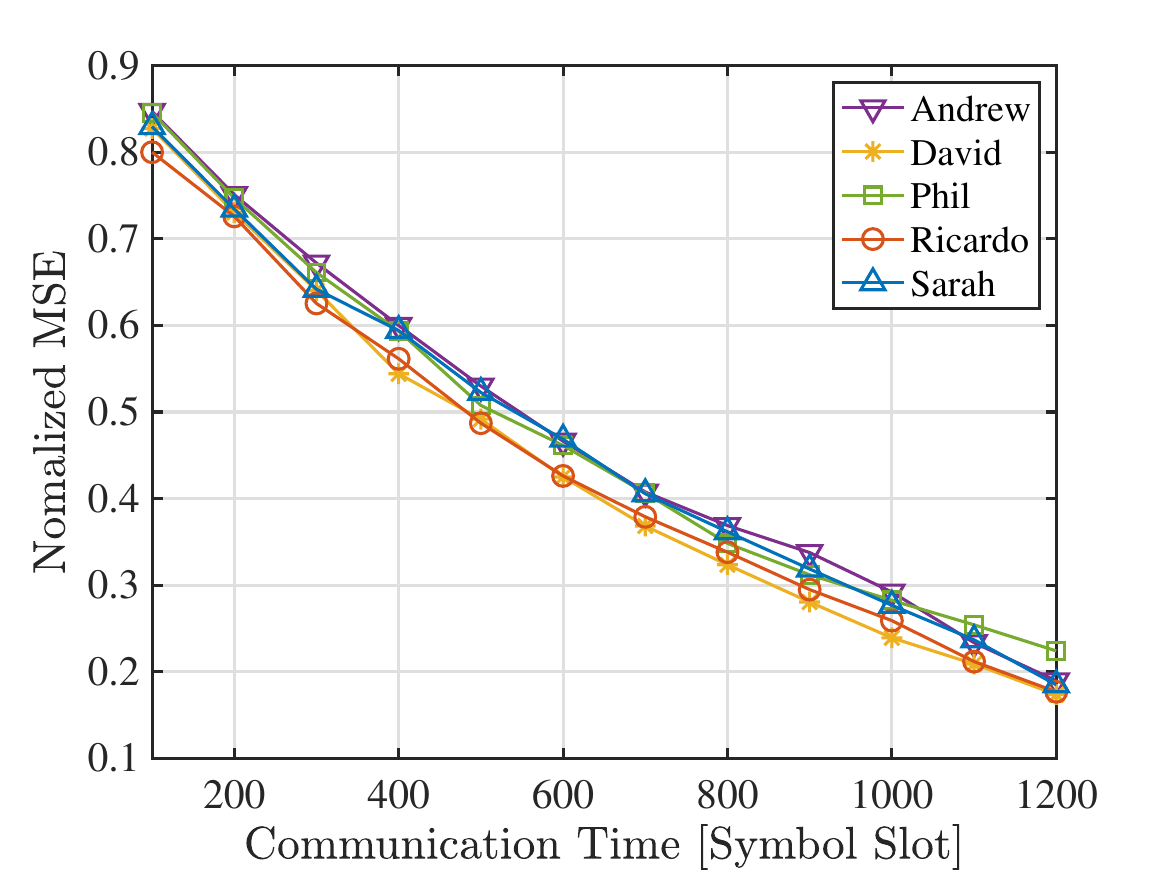}
    \label{subfig:all}
    }
    \subfigure[]{\includegraphics[width=0.45\textwidth]{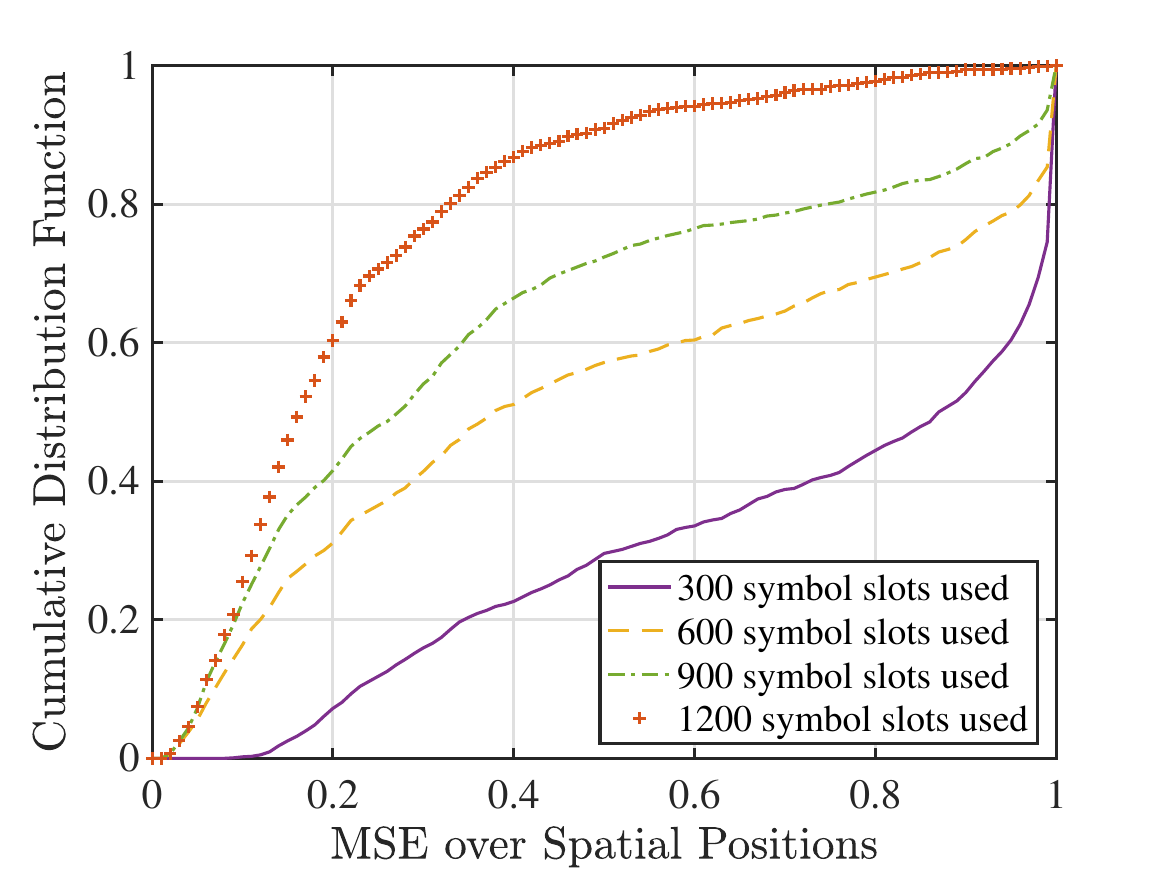}
    \label{subfig:error_dist}
    }
    \caption{Error evaluation of the proposed FlyCom$^2$.}
    \label{fig:progressive}
\end{figure}
First, we verify the progressive performance improvement of distributed PtCloud fusion achieved by the proposed FlyCom$^2$ in Fig.~\ref{fig:progressive}. The fusion performance is measured by the MSE between the prediction results using~\eqref{eq:global_predictor} and their ground truth, as defined in~\eqref{eq:MSPE} and within different PtCloud datasets, the MSE is normalized by their corresponding variances in~\eqref{eq:GPRjoint_dist}. We focus on the progressive PtCloud fusion over all depth-6 voxels generated along with the octree space search, where communication symbol slots from $200$ to $1200$. Furthermore, the prediction areas in~\eqref{eq:global_predictor} are given as spatial coordinates uniformly sampled on the shape reconstructed using the occupancy feedback during the depth-6 octree space search. The curves of the MSE versus the communication time are presented in Fig.~\ref{subfig:all}, from which one can observe that the fusion error keeps decreasing as the FlyCom$^2$ progresses. The trend remains unchanged for all five testing datasets. This is due to the continuous observation accumulation at the server for the global fusion. This conclusion is additionally confirmed by the results in Fig.~\ref{subfig:error_dist}, which shows that the error distribution for individual sampled spatial positions is continuously improved as the symbol slots used in FlyCom$^2$ changes from $300$ to $1200$.

\begin{figure}[t]
    \centering
    \subfigure[Transmit SNR $\gamma = 5$ dB]{\includegraphics[width=0.45\textwidth]{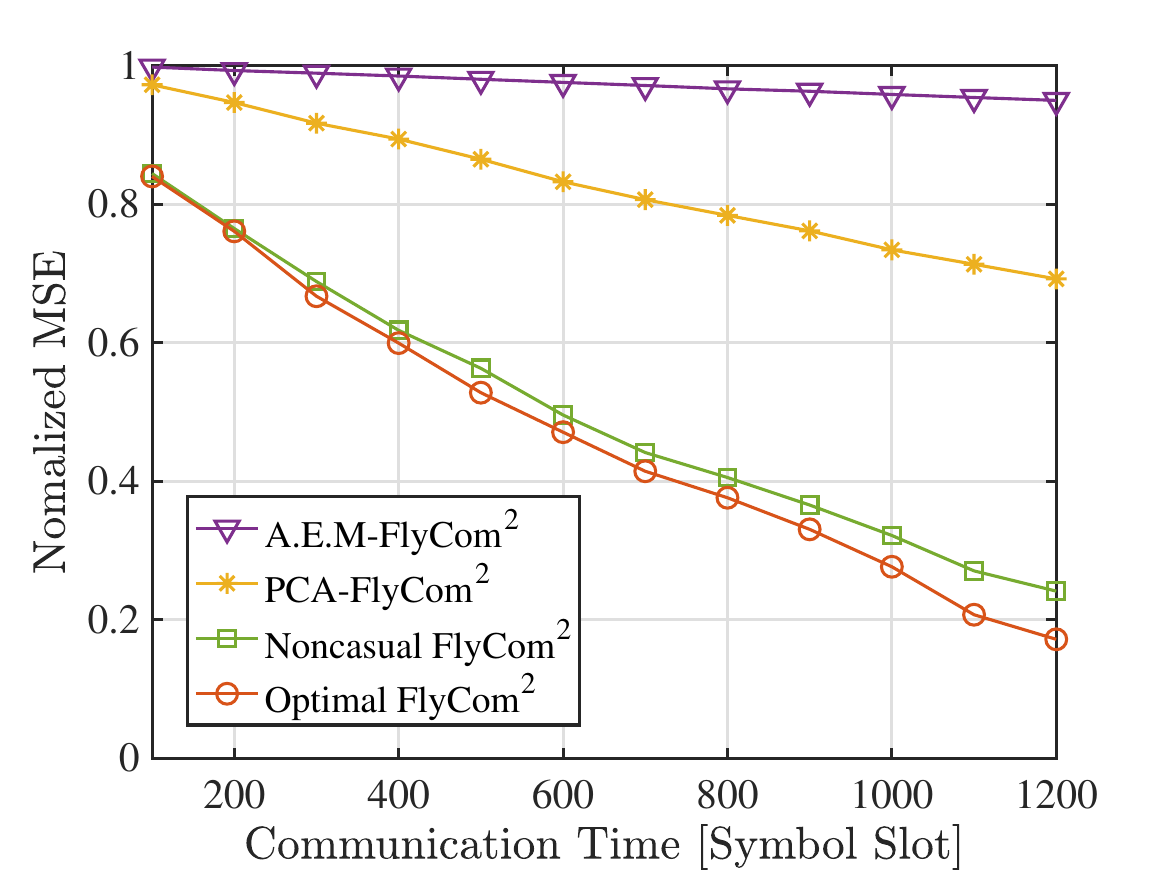}
    \label{subfig:nmspe_5dB}
    }
    \subfigure[Transmit SNR $\gamma = 15$ dB]{\includegraphics[width=0.45\textwidth]{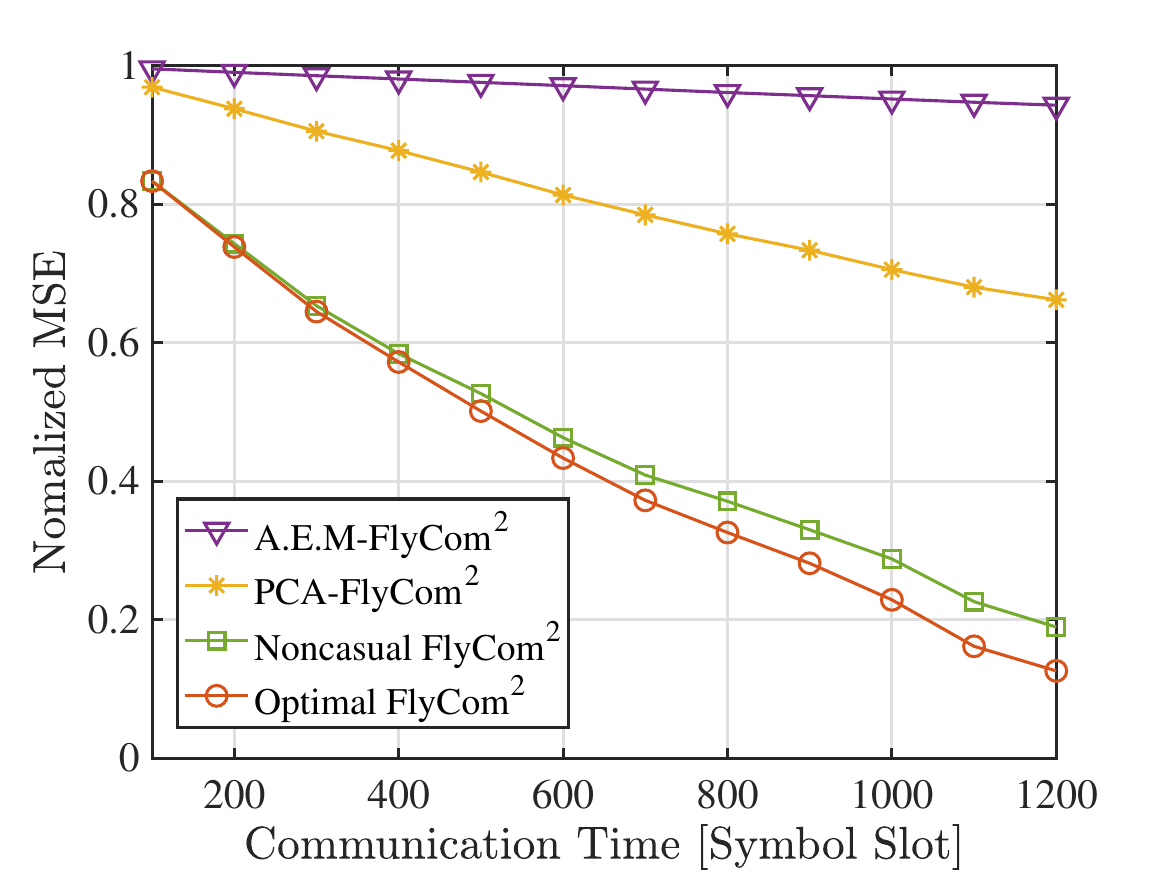}
    \label{subfig:nmspe_15dB}
    }
    \caption{Performance comparison between the proposed FlyCom$^2$ method and benchmarking schemes within different FlyCom$^2$ progresses.}
    \label{fig:diff_T}
\end{figure}
\begin{figure}[t]
    \centering
    \includegraphics[width=0.48\textwidth]{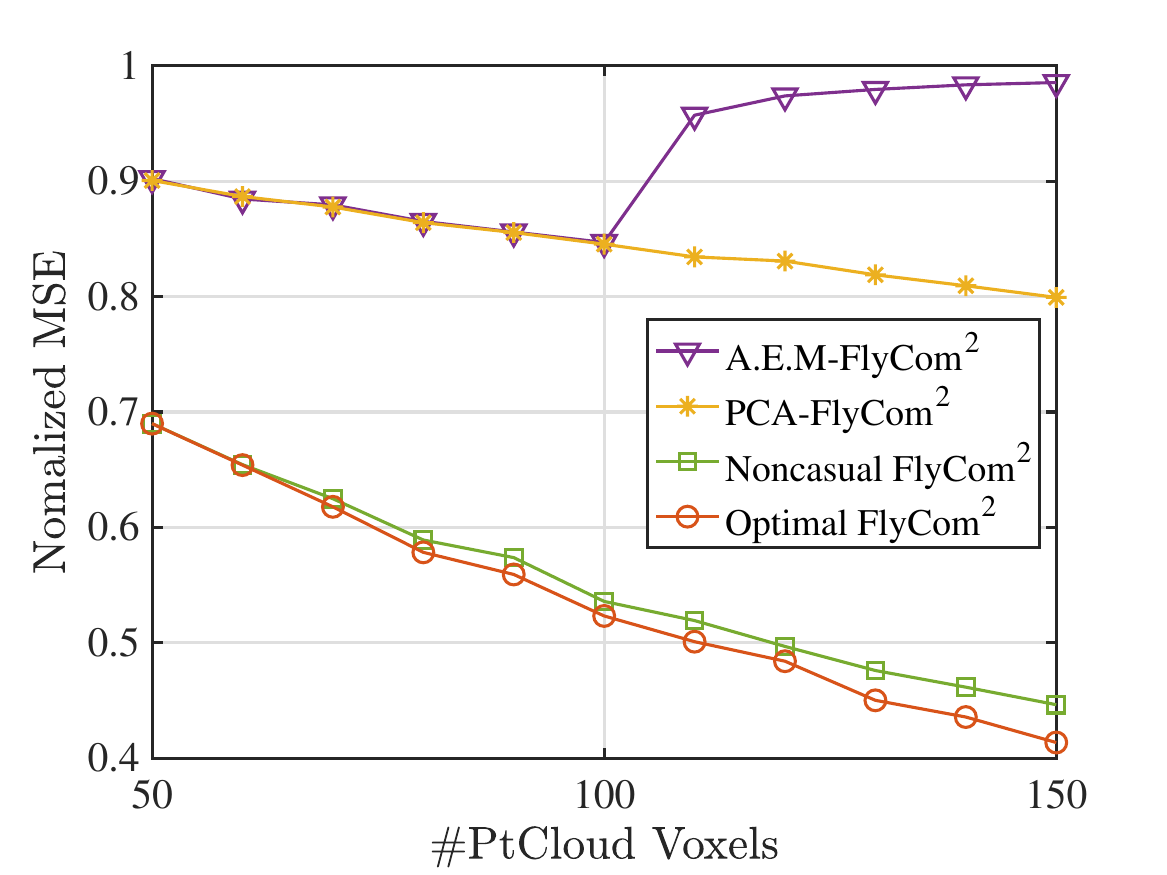}
    \caption{Performance comparison between the proposed FlyCom$^2$ method and benchmarking schemes within different PtCloud volumes processed per slot.}
    \label{fig:diff_N}
\end{figure}

Then, we compare the error performance of the proposed FlyCom$^2$ with that of the benchmarking schemes, varying the communication time and the PtCloud volume processed per slot in Fig.~\ref{fig:diff_T} and Fig.~\ref{fig:diff_N} respectively. The progressive fusion settings are the same as these in Fig.~\ref{fig:progressive}. The curves of the fusion error versus the communication time are plotted in Fig.~\ref{fig:diff_T} with the transmit SNR given as $5$ and $15$ dB. It is observed from Fig.~\ref{fig:diff_T} that all schemes achieve an error reduction as the progressive fusion progresses. Compared to PCA-FlyCom$^2$, both noncasual FlyCom$^2$ and the optimal FlyCom$^2$ can realize error reduction with a significantly faster rate. This is attributed to their joint optimization of the observation matrix and AirComp which balances the tradeoff between the data heterogeneity and AirComp error suppression (see Remark~\ref{Remark:tradeoff}). Primed with noncasual FlyCom$^2$, the optimal FlyCom$^2$ further takes into account the temporal correlation between the progressive operations (see Remark~\ref{Remark:uncorrelated}) in the observation design, leading to a further performance improvement, as reflected by the approximate $25\%$ error reduction within the consumed communication symbol slots of $1200$ in Fig.~\ref{subfig:nmspe_5dB}. Besides, we observe that the performance gap becomes larger as the communication time increases, because of the enlarged possibility of current PtCloud samples showing a strong correlation with the accumulated observations. Besides, Fig.~\ref{fig:diff_T} shows that the A.E.M.-FlyCom$^2$ realizes the lowest error reduction, which is consistent with our conclusion in Remark~\ref{Remark:tradeoff} that minimizing the aggregation error by selecting non-principal components in local observations will constrain the sensing knowledge retained for the global prediction.

The error performance achieved by different schemes within the number of PtClod voxels processed per slot changing from $50$ to $150$ is shown in Fig.~\ref{fig:diff_N}, where the communication time is fixed as $600$ symbol slots. We observe from Fig.~\ref{fig:diff_N} that increasing the voxel population can lead to error reduction in PCA-FlyCom$^2$, noncasual FlyCom$^2$ and the optimal FlyCom$^2$. This is because more PtCloud information is captured by the local observations in this process to benefit the global fusion. At the same time, the optimal FlyCom$^2$ outperforms noncasual FlyCom$^2$ and increasing the PtCloud population enlarges this performance gap due to the resulting larger temporal correlation. Compared to the above schemes, PCA-FlyCom$^2$ still reveals a much slower error reduction because of its lack of controlling the multi-sensor heterogeneity and channel diversity as discussed in Remark~\ref{Remark:tradeoff}. Besides, it is observed that when the number of PtCloud voxels does not exceed $100$, A.E.M.-FlyCom$^2$ achieves the same performance as PCA-FlyCom$^2$. This is straightforwardly due to that, in this region, the observation matrix dimensionality is smaller than or equal to the AirComp symbol length $I=100$, meaning that the principal components selected in PCA-FlyCom$^2$ overlap the non-principal components selected in A.E.M.-FlyCom$^2$. Further enlarging the voxel population will lead to A.E.M.-FlyCom$^2$ giving up the relatively useful principal components of on-sensor PtClouds, resulting in a performance degradation in the global fusion. This is verified by the error increase of A.E.M.-FlyCom$^2$ after the voxel population being larger than $100$ in Fig.~\ref{fig:diff_N}. 

\subsection{PtCloud Rendering}
\begin{figure*}
     \centering
    \subfigure[Oracle fusion, ]{\includegraphics[width=0.23\textwidth]{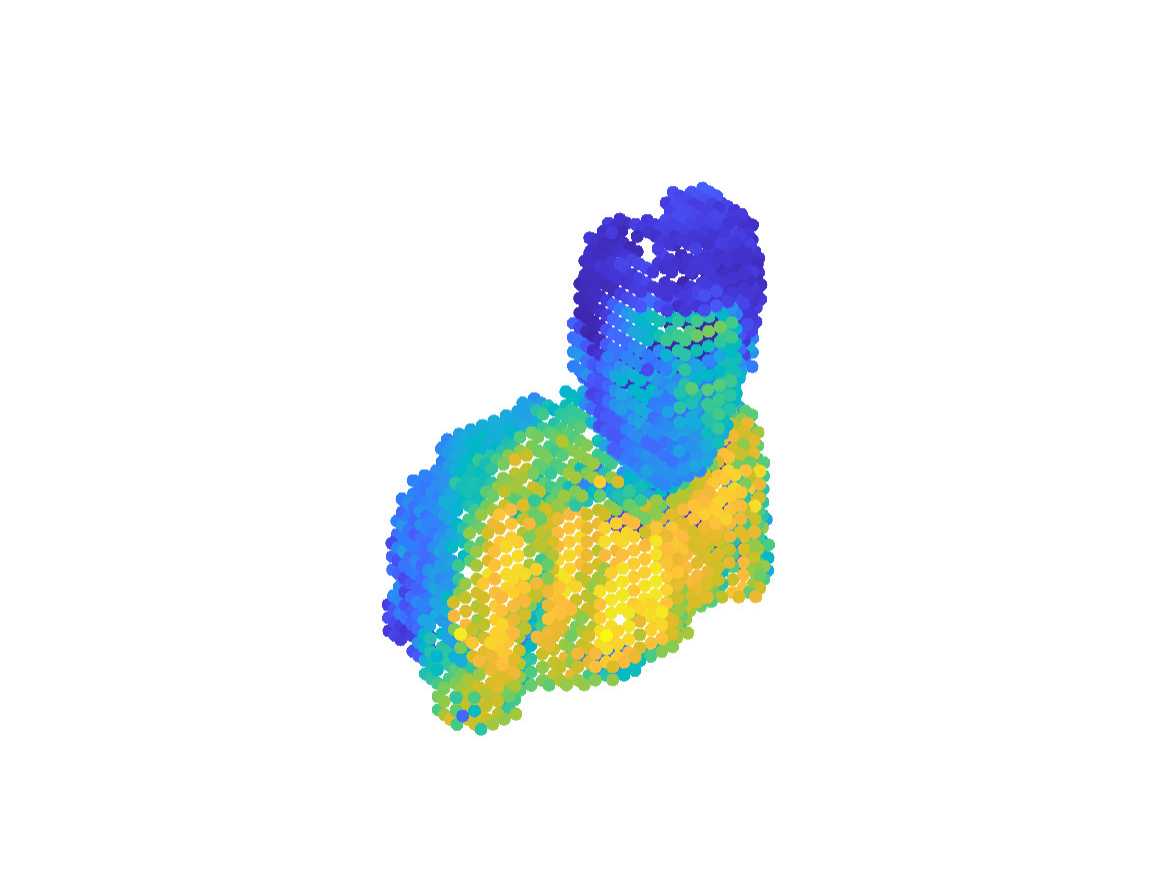}
        \label{subfig:perfect}
    }    
    \subfigure[$33\%$ voxels proceed]{\includegraphics[width=0.23\textwidth]{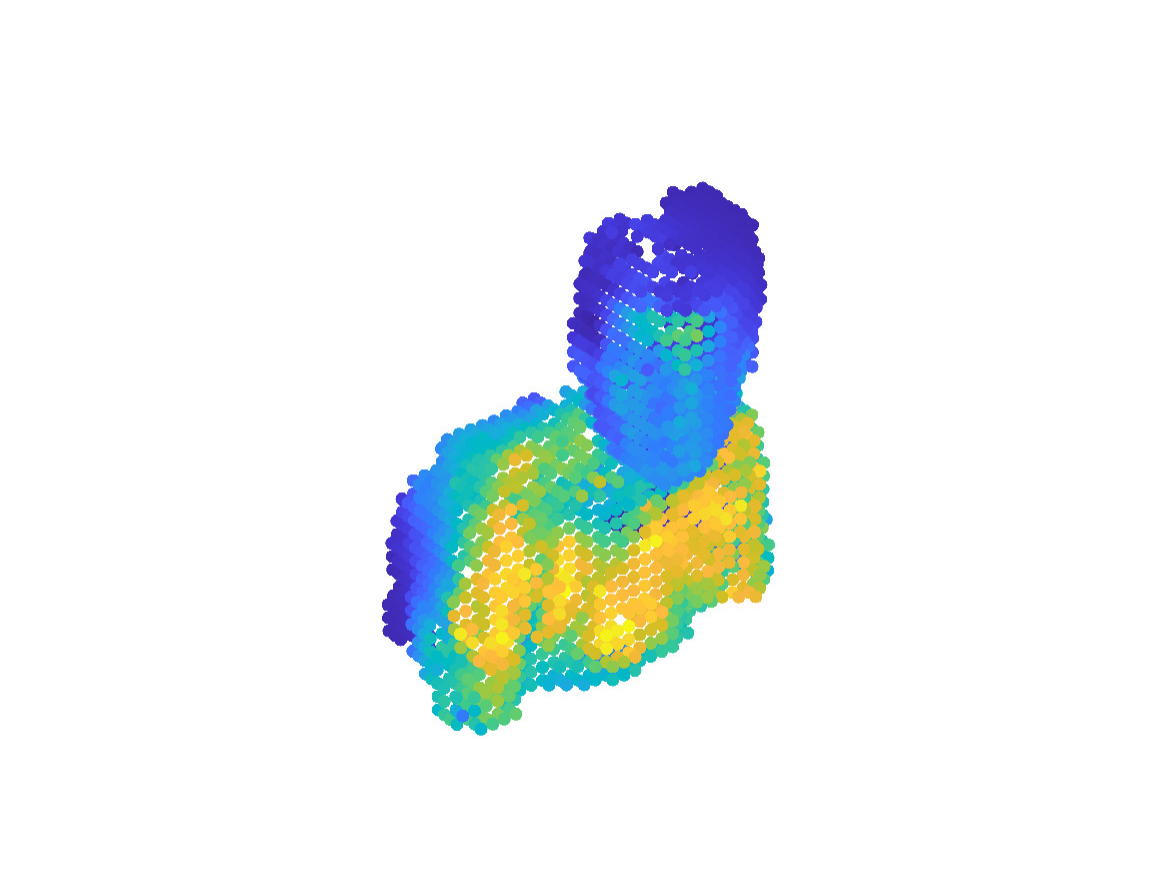}
    \label{subfig:andrew1_3}
    }
    \subfigure[$66\%$ voxels proceed]{\includegraphics[width=0.23\textwidth]{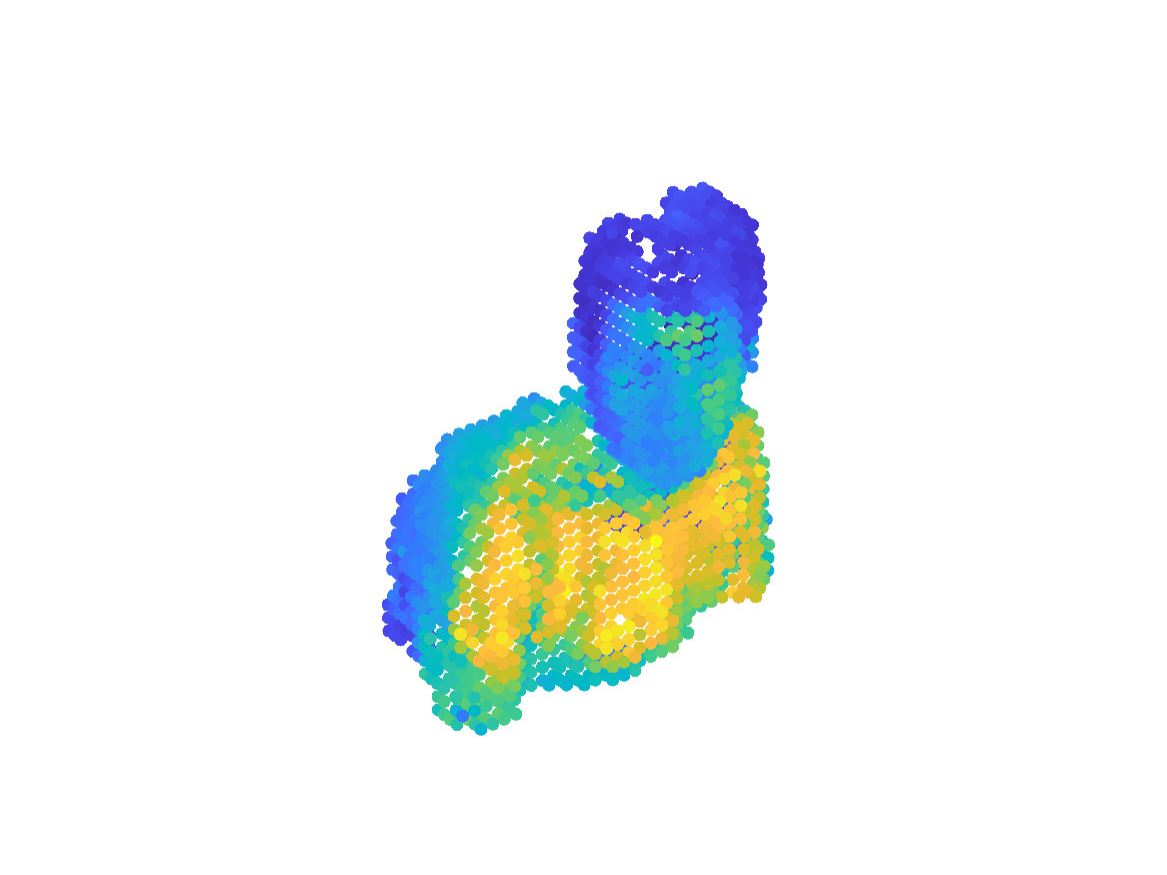}
    \label{subfig:andrew2_3}
    }
    \subfigure[$99\%$ voxels proceed]{\includegraphics[width=0.23\textwidth]{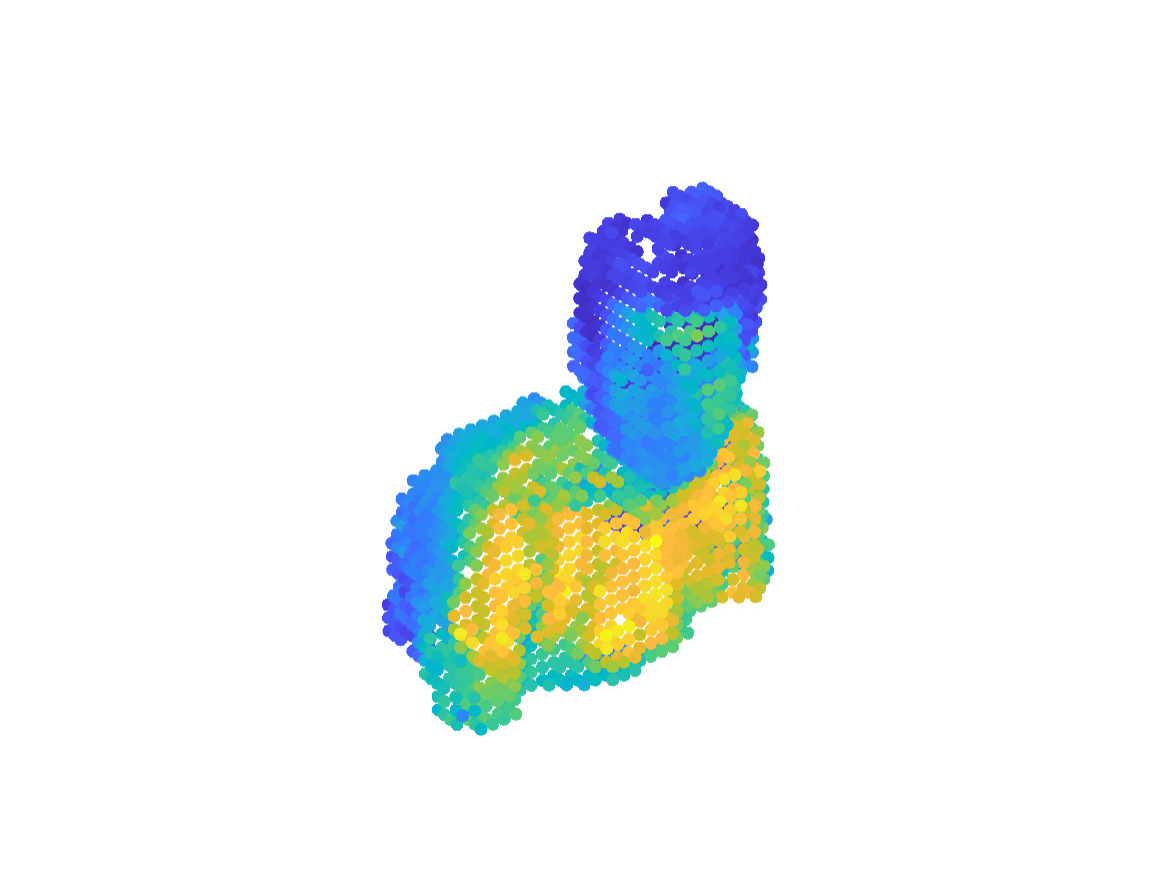}
    \label{subfig:andrew3_3}
    }
    \caption{PtCloud rendering of \emph{Andrew} within the distributed PtCloud fusion over depth $d=5$ voxels, using the Oracle aggregation in (a) and the FlyCom$^2$ method in (b), (c), (d) with different progresses.}
    \label{fig:rendering_andrew_d5}
\end{figure*}
\begin{figure*}
     \centering
    \subfigure[Oracle fusion]{\includegraphics[width=0.23\textwidth]{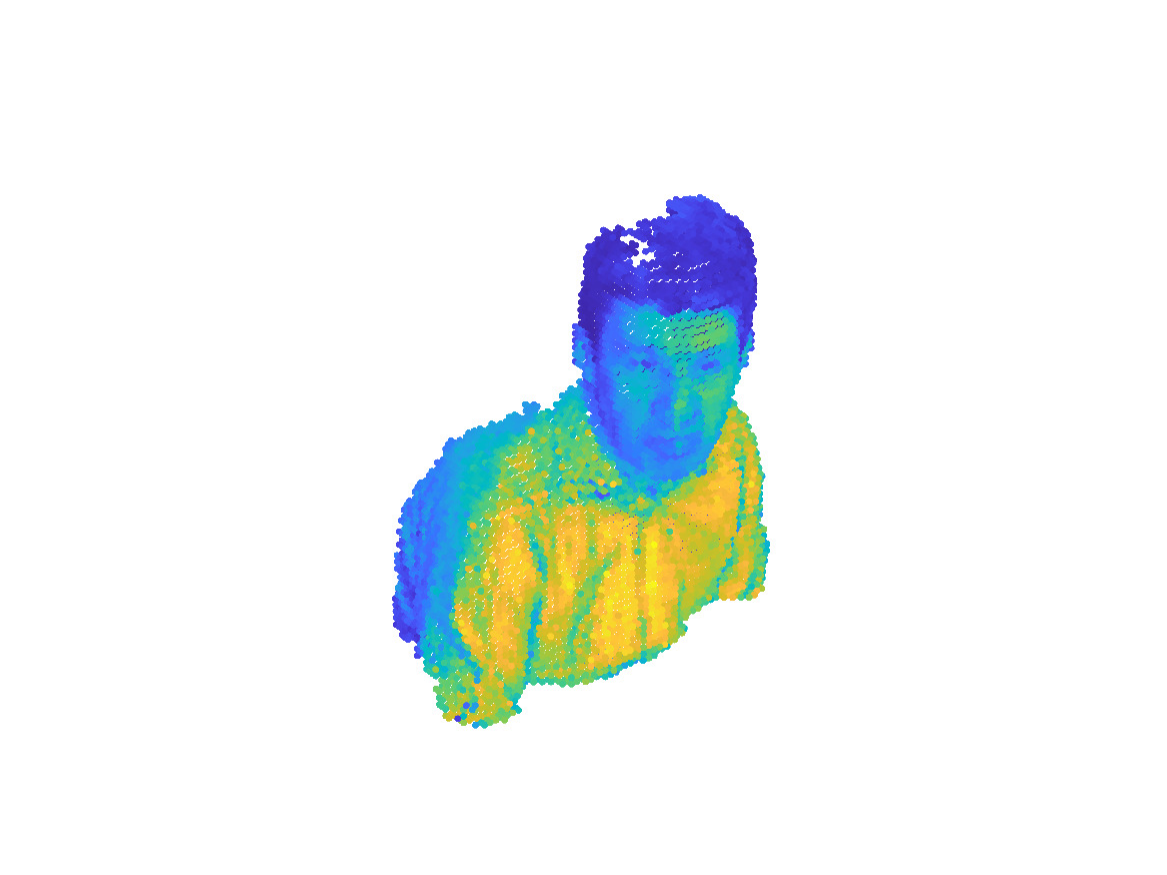}
    \label{subfig:perfect}
    }    
    \subfigure[$33\%$ voxels proceed]{\includegraphics[width=0.23\textwidth]{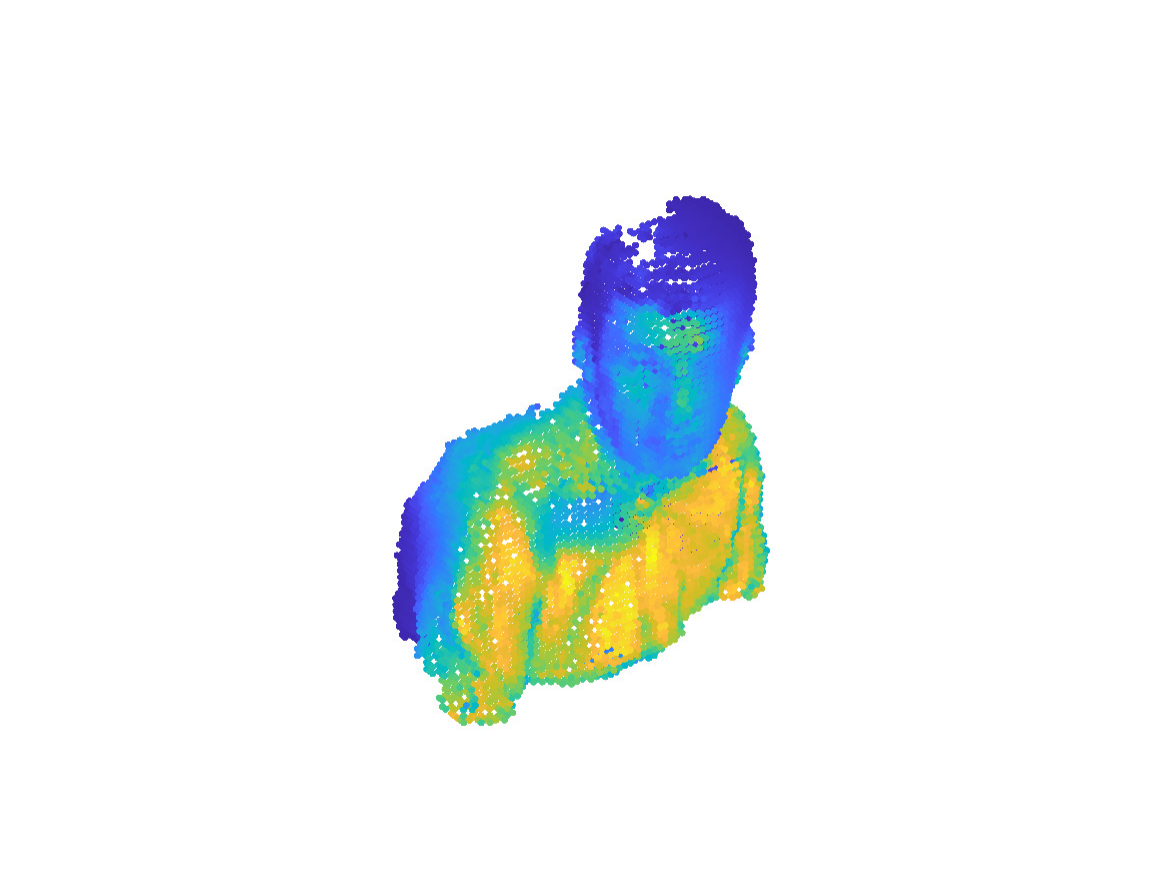}
    \label{subfig:andrew1_3}
    }
    \subfigure[$66\%$ voxels proceed]{\includegraphics[width=0.23\textwidth]{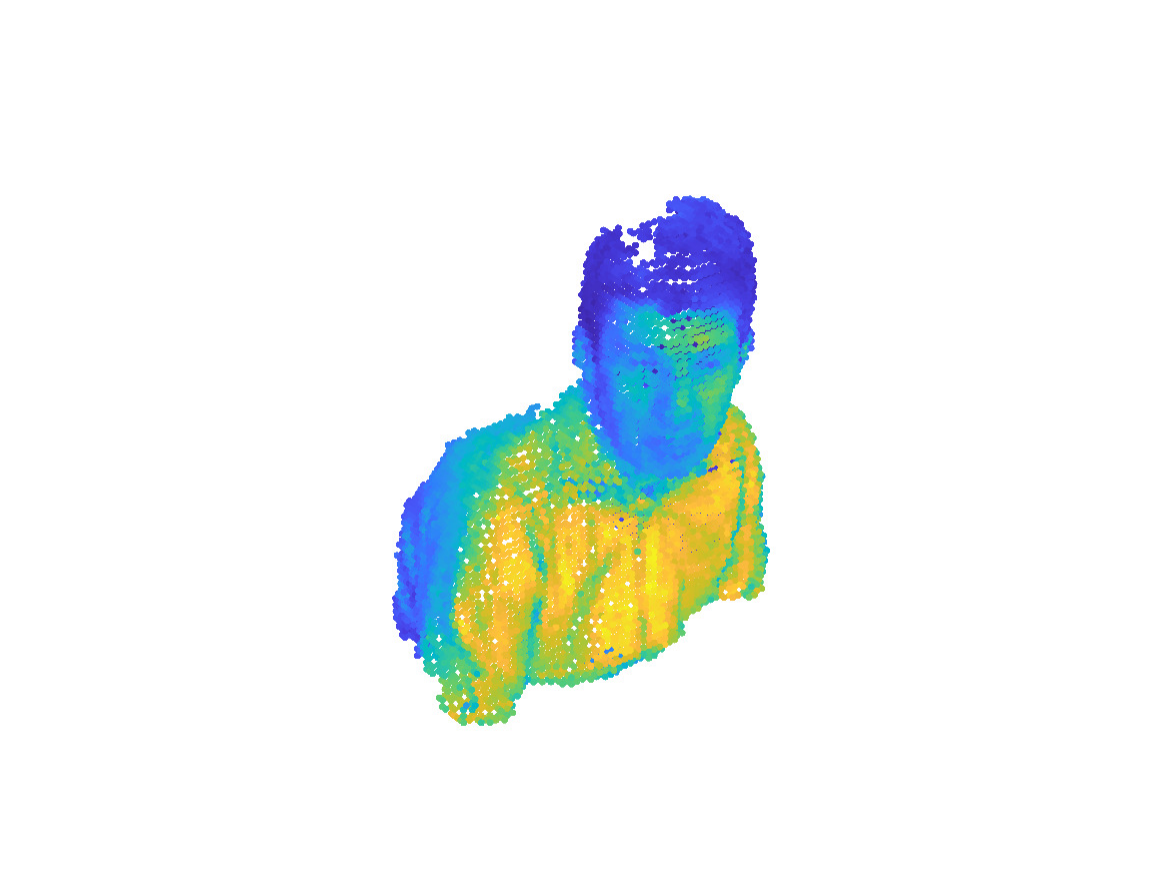}
    \label{subfig:andrew2_3}
    }
    \subfigure[$99\%$ voxels proceed]{\includegraphics[width=0.23\textwidth]{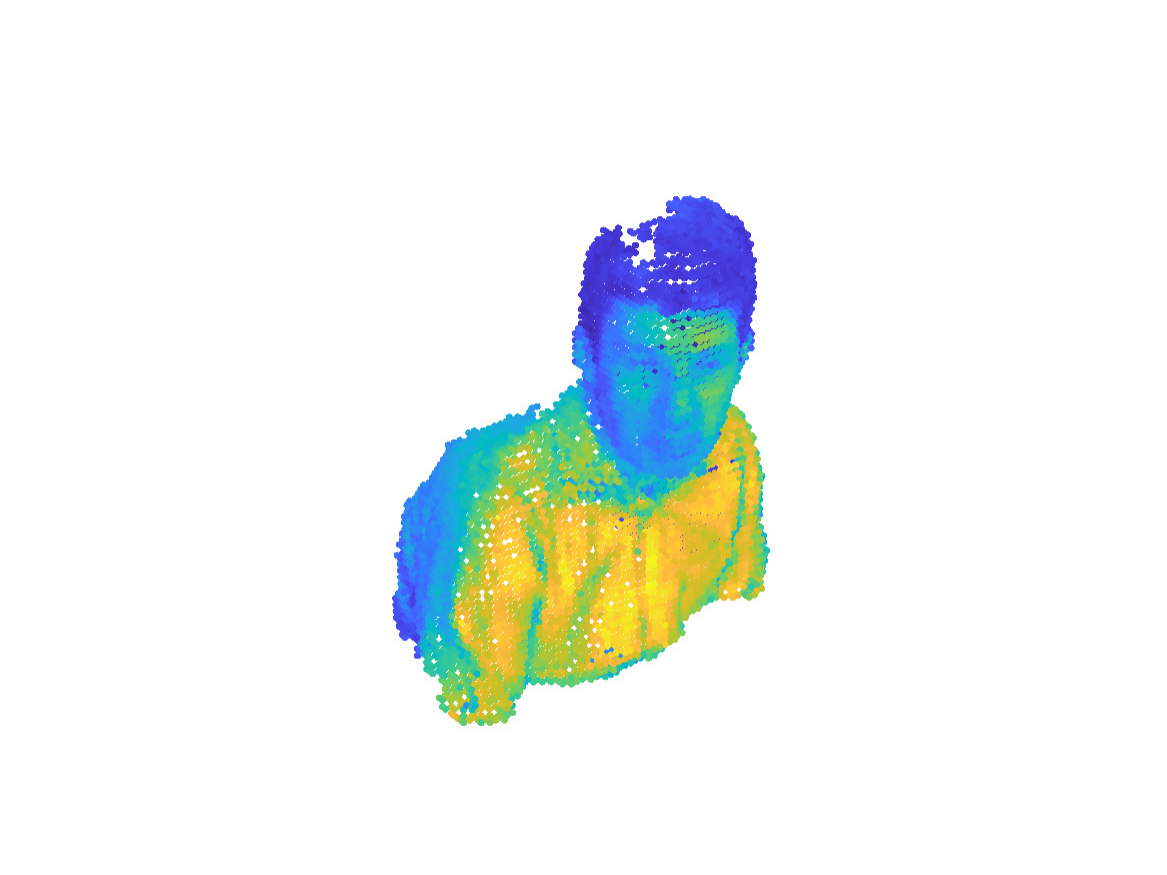}
    \label{subfig:andrew3_3}
    }
    \caption{PtCloud rendering of \emph{Andrew} within the distributed PtCloud fusion over depth $d=6$ voxels, using the Oracle aggregation in (a) and the FlyCom$^2$ method in (b), (c), (d) with different progresses}
    \label{fig:rendering_andrew_d6}
\end{figure*}

We also compare the PtCloud rendering result using the proposed FlyCom$^2$ method with that of using the Oracle fusion in Fig.~\ref{fig:rendering_andrew_d5} and Fig.~\ref{fig:rendering_andrew_d6}, based on the PtCloud data of \emph{Andrew} with voxelization depth of $5$ and $6$, respectively. The spatial positions are uniformly sampled from the shape reconstructed within the corresponding depth.
As one can see, with the stepping of the FlyCom$^2$-based distributed PtCloud fusion, the scene rendered at the server is refined with more point details due to the sequential collection of informative fused PtCloud sequences.
For example, the face of Andrew presents more details, including the emergence of eyeballs at the $66\%$ voxels proceed compared with the $33\%$ one at a depth of $6$. After all the FlyCom$^2$ steps, the rendering results are quite similar to the Oracle fusion ones, demonstrating the superiority of the proposed FlyCom$^2$ distributed PtCloud fusion, which achieves a comparable distributed PtCloud fusion performance while significantly dropping the PtCloud uploading cost versus the rigid PtCloud fusion schemes using raw data uploading.

\section{Conclusions}\label{section:conclusions}
In this work, we present the FlyCom$^2$ framework for ISEA to address the significant challenges of computational complexity and communication bottlenecks in distributed PtCloud fusion. The proposed FlyCom$^2$ facilitates efficient and progressive PtCloud aggregation by aligning the fusion process with Gaussian process regression, ensuring continuous improvement in global PtCloud representation and offering a robust solution to the limitations of existing one-shot compression and aggregation methods. Furthermore, we completed the joint optimization of local observation synthesis and AirComp receiver settings, addressing the temporal correlations between progressive communication-and-computing operations and the tradeoff between communication distortions and data heterogeneity.

Future research can further refine the FlyCom$^2$ framework by exploring advanced machine learning models and algorithms tailored for PtCloud processing and fusion. Investigating the integration of FlyCom$^2$ with emerging 6G technologies, such as terahertz communication and quantum computing, could unlock new levels of performance and efficiency. Additionally, developing standardized protocols and interfaces for seamless interoperability between diverse sensors and AI models will be crucial for widespread adoption.

\appendix
\subsection{Proof of Lemma~\ref{Lemma:monotonicity}}\label{Apdx:monotonicity}
Given positive-definite matrices $\mathbf{M}_1$, $\mathbf{M}_2$ and $\mathbf{M}_1\succeq\mathbf{M}_2$, there is $\mathbf{\Delta}=\mathbf{M}_1-\mathbf{M}_2$ is a positive semidefinite matrix. Then, we have
\begin{equation*}
\begin{aligned}
        \mathsf{Tr}\left(\frac{\mathbf{M}_0}{\mathbf{M}_1}\right) &= \mathsf{Tr}\left(\frac{\mathbf{M}_0}{\mathbf{M}_2 + \mathbf{M}_1-\mathbf{M}_2}\right)\\
    &=\mathsf{Tr}\left(\frac{\mathbf{M}_0}{\mathbf{M}_2 + \mathbf{U}_{\mathbf{\Delta}}\mathbf{\Sigma}_{\mathbf{\Delta}}\mathbf{U}_{\mathbf{\Delta}}^{\top}}\right),
\end{aligned}
\end{equation*}
where the thin eigenvalue decomposition $\mathbf{\Delta}=\mathbf{U}_{\mathbf{\Delta}}\mathbf{\Sigma}_{\mathbf{\Delta}}\mathbf{U}_{\mathbf{\Delta}}^{\top}$ is used. Using the Woodbury matrix identity~\cite{MatrixAnalysis}, the inverse of the sum matrix $(\mathbf{M}_2 + \mathbf{U}_{\mathbf{\Delta}}\mathbf{\Sigma}_{\mathbf{\Delta}}\mathbf{U}_{\mathbf{\Delta}}^{\top})^{-1}$ can be rewritten as
\begin{equation*}
    \begin{aligned}
        &(\mathbf{M}_2 + \mathbf{U}_{\mathbf{\Delta}}\mathbf{\Sigma}_{\mathbf{\Delta}}\mathbf{U}_{\mathbf{\Delta}}^{\top})^{-1}\\
        &= \mathbf{M}_2^{-1} - \mathbf{M}_2^{-1}\mathbf{U}_{\mathbf{\Delta}}\left(\mathbf{\Sigma}_{\mathbf{\Delta}}^{-1}+\mathbf{U}_{\mathbf{\Delta}}^{\top}\mathbf{M}_2^{-1}\mathbf{U}_{\mathbf{\Delta}}\right)^{-1}\mathbf{U}_{\mathbf{\Delta}}^{\top}\mathbf{M}_2^{-1}\\
        & \overset{\triangle}{=} \mathbf{M}_2^{-1} - \mathbf{\Delta}^{\prime},
    \end{aligned}
\end{equation*}
which leads to 
\begin{equation*}
 \mathsf{Tr}\left(\frac{\mathbf{M}_0}{\mathbf{M}_1}\right) = \mathsf{Tr}\left(\frac{\mathbf{M}_0}{\mathbf{M}_2}\right)-\mathsf{Tr}\left(\mathbf{\Delta}^{\prime}\mathbf{M}_0\right).
\end{equation*}
Next, let's prove that $\mathbf{\Delta}^{\prime}$ is a positive-semidefinite matrix. Specifically, since $\mathbf{M}_2\succ \mathsf{0}$ and $\mathbf{\Delta}\succeq \mathsf{0}$, there is $\mathbf{\Sigma}_{\mathbf{\Delta}}^{-1}\succ \mathsf{0}$ and $\mathbf{U}_{\mathbf{\Delta}}^{\top}\mathbf{M}_2^{-1}\mathbf{U}_{\mathbf{\Delta}}\succ \mathsf{0}$, eventually giving
\begin{equation*}
\left(\mathbf{\Sigma}_{\mathbf{\Delta}}^{-1}+\mathbf{U}_{\mathbf{\Delta}}^{\top}\mathbf{M}_2^{-1}\mathbf{U}_{\mathbf{\Delta}}\right)^{-1}\succ \mathsf{0}.
\end{equation*}
Therefore, for any vector $\mathbf{x}$, let $y=\mathbf{U}_{\mathbf{\Delta}}^{\top}\mathbf{M}_2^{-1}\mathbf{x}$ and we have
\begin{equation*}
    \begin{aligned}
        \mathbf{x}^{\top}\mathbf{\Delta}^{\prime}\mathbf{x}&=\mathbf{y}^{\top}\left(\mathbf{\Sigma}_{\mathbf{\Delta}}^{-1}+\mathbf{U}_{\mathbf{\Delta}}^{\top}\mathbf{M}_2^{-1}\mathbf{U}_{\mathbf{\Delta}}\right)^{-1}\mathbf{y}\\
        &\geq 0,
    \end{aligned}
\end{equation*}
which suggests that $\mathbf{\Delta}^{\prime}\succeq\mathbf{0}$. Finally, as both $\mathbf{\Delta}^{\prime}$ and $\mathbf{M}_0$ are positive-semidefinite, using the Ruhe’s trace inequality gives that 
\begin{equation*}
    \mathsf{Tr}\left(\mathbf{\Delta}^{\prime}\mathbf{M}_0\right) \geq \sum_{n=1}^N\lambda_{\mathbf{\Delta}^{\prime},n}\lambda_{\mathbf{M}_0,N-n+1}\geq0,
\end{equation*}
where $\lambda_{*,n}$ denotes the $n$-th largest eigenvalue of the matrix indicated. Hence, we can get $\mathsf{Tr}\left(\frac{\mathbf{M}_0}{\mathbf{M}_1}\right) \leq\mathsf{Tr}\left(\frac{\mathbf{M}_0}{\mathbf{M}_2}\right)$.
\subsection{Proof of Lemma~\ref{Lemma:bound}}\label{Apdx:matrix_bound}
For ease of notation, let's define the eigenvalues of $\mathbf{\Psi}_k$ as $\{\lambda_{\mathbf{\Psi}_k,n}\}_{1\leq n\leq N}$, $\bar{\lambda}_{\mathbf{\Psi}_k} = \frac{1}{N}\sum_n \lambda_{\mathbf{\Psi}_k,n}$, and $k^* = \arg\max_k\  \bar{\lambda}_{\mathbf{\Psi}_k}$. Then, the Lemma can be proved as follows.
\subsubsection{Left Inequality}
Using the definition of $\mathbf{\Psi}$ and $\bar{\mathbf{W}}$, $\bar{\mathbf{W}}^{\top}\mathbf{\Psi}\bar{\mathbf{W}}$ can be written as
\begin{equation*}
    \begin{aligned}
        \frac{1}{\kappa}\bar{\mathbf{W}}^{\top}\mathbf{\Psi}\bar{\mathbf{W}} &=\frac{1}{\kappa N}\sum_k\mathbf{\Psi}_{k^*}\\
        & = \frac{1}{\kappa N}\mathbf{\Psi}_{k^*} + \frac{1}{\kappa N}\sum_{k=1,k\neq k^*}^K\mathbf{\Psi}_{k},
    \end{aligned}
\end{equation*}
where given $\kappa$ defined in the Lemma statement, the minimum eigenvalue of the first term $\frac{1}{\kappa N}\mathbf{\Psi}_{k^*}$ is larger than $\bar{\lambda}_{\mathbf{\Psi}_{k^*}}$, leading to 
\begin{equation*}
    \frac{1}{\kappa N}\mathbf{\Psi}_{k^*}\succeq \bar{\lambda}_{\mathbf{\Psi}_{k^*}} \mathbf{I}_N.
\end{equation*} Furthermore, the second Hermitian matrix $\frac{1}{\kappa N}\sum_{k=1,k\neq k^*}^K\mathbf{\Psi}_{k}$ has nonnegative eigenvalues, eventually resulting in 
\begin{equation*}
    \frac{1}{\kappa}\bar{\mathbf{W}}^{\top}\mathbf{\Psi}\bar{\mathbf{W}}\succeq \bar{\lambda}_{\mathbf{\Psi}_{k^*}} \mathbf{I}_N.
\end{equation*}

\subsubsection{Right Inequality} On the other hand, since $k^* = \arg\max_k\  \bar{\lambda}_{\mathbf{\Psi}_k}$, there is
\begin{equation*}
    \bar{\lambda}_{\mathbf{\Psi}_{k^*}}\geq \frac{1}{K}\sum_k\bar{\lambda}_{\mathbf{\Psi}_k} = \frac{1}{K}\mathsf{Tr}\left(\bar{\mathbf{W}}^{\top}\mathbf{\Psi}\bar{\mathbf{W}}\right).
\end{equation*}
Furthermore, $\bar{\mathbf{W}}^{\top}\mathbf{\Psi}\bar{\mathbf{W}}$ is a positive-definite matrix, namely its eigenvalues are positive. Hence, the trace of $\bar{\mathbf{W}}^{\top}\mathbf{\Psi}\bar{\mathbf{W}}$, being equal to the summation of its eigenvalues, shall be larger than each eigenvalue, leading to $\mathsf{Tr}\left(\bar{\mathbf{W}}^{\top}\mathbf{\Psi}\bar{\mathbf{W}}\right)\mathbf{I}_N- \bar{\mathbf{W}}^{\top}\mathbf{\Psi}\bar{\mathbf{W}}$ always being positive semidefinite. Therefore, there is
\begin{equation*}
    \bar{\lambda}_{\mathbf{\Psi}_{k^*}} \mathbf{I}_N\succeq \frac{1}{K}\mathsf{Tr}\left(\bar{\mathbf{W}}^{\top}\mathbf{\Psi}\bar{\mathbf{W}}\right)\mathbf{I}_N\succeq \frac{1}{K}\bar{\mathbf{W}}^{\top}\mathbf{\Psi}\bar{\mathbf{W}},
\end{equation*}
which completes the proof.
\subsection{Proof of Lemma~\ref{Lemma:identity}}\label{Apdx:identity}
First, using the property of block matrix inverse~\cite{MatrixAnalysis}, there is 
\begin{equation*}
\begin{aligned}
    &\left(    \begin{bmatrix}
        \mathbf{I}&\\
        &\mathbf{W}^{\top}
    \end{bmatrix}\begin{bmatrix}
        \mathbf{P}_1&\mathbf{P}_2\\
        \mathbf{P}_2^{\top}&\mathbf{P}_3
    \end{bmatrix}\begin{bmatrix}
        \mathbf{I}&\\
        &\mathbf{W}
    \end{bmatrix}\right)^{-1} \\
    &=\begin{bmatrix}
        \mathbf{P}_1&\mathbf{P}_2\mathbf{W}\\
        \mathbf{W}^{\top}\mathbf{P}_2^{\top}&\mathbf{W}^{\top}\mathbf{P}_3\mathbf{W}
    \end{bmatrix}^{-1}\\
    & = \begin{bmatrix}
        \mathbf{P}_1^{-1} + \mathbf{P}_1^{-1}\mathbf{P}_2\mathbf{W}\mathbf{D}^{-1}\mathbf{W}^{\top}\mathbf{P}_2^{\top}\mathbf{P}_1^{-1}&-\mathbf{P}_1^{-1}\mathbf{P}_2\mathbf{W}\mathbf{D}^{-1}\\
        -\mathbf{D}^{-1}\mathbf{W}^{\top}\mathbf{P}_2^{\top}\mathbf{P}_1^{-1}&\mathbf{D}^{-1}
    \end{bmatrix}\\
    &\overset{\triangle}{=}\mathbf{A},
\end{aligned} 
\end{equation*}
with $\mathbf{D} = \mathbf{W}^{\top}\mathbf{P}_3\mathbf{W} - \mathbf{W}^{\top}\mathbf{P}_2^{\top}\mathbf{P}_1^{-1}\mathbf{P}_2\mathbf{W} = \mathbf{W}^{\top}(\mathbf{P}_3-\mathbf{\Upsilon})\mathbf{W}$ with $\mathbf{\Upsilon}=\mathbf{P}_2^{\top}\mathbf{P}_1^{-1}\mathbf{P}_2$. Then, the ratio trace in Lemma~\ref{Lemma:identity} can be rewritten as 
\begin{equation*}
    \begin{aligned}
    \mathsf{RT} \overset{\triangle}{=}& \mathsf{Tr}\left(\mathbf{A}\begin{bmatrix}
        \mathbf{I}&\\
        &\mathbf{W}^{\top}
    \end{bmatrix}\begin{bmatrix}
        \mathbf{Q}_1&\mathbf{Q}_2\\
        \mathbf{Q}_2^{\top}&\mathbf{Q}_3
    \end{bmatrix}\begin{bmatrix}
        \mathbf{I}&\\
        &\mathbf{W}
    \end{bmatrix}\right)\\
    =& \mathsf{Tr}\left(\mathbf{P}_1^{-1}\mathbf{Q}_1 + \mathbf{P}_1^{-1}\mathbf{P}_2\mathbf{W}\mathbf{D}^{-1}\mathbf{W}^{\top}\mathbf{P}_2^{\top}\mathbf{P}_1^{-1}\mathbf{Q}_1\right)\\
    &+\mathsf{Tr}\left(\mathbf{D}^{-1}\mathbf{W}^{\top}\mathbf{Q}_3\mathbf{W}\right)-\mathsf{Tr}\left(\mathbf{D}^{-1}\mathbf{W}^{\top}\mathbf{P}_2^{\top}\mathbf{P}_1^{-1}\mathbf{Q}_2\mathbf{W} \right) \\
    &-\mathsf{Tr}\left(\mathbf{P}_1^{-1}\mathbf{P}_2\mathbf{W}\mathbf{D}^{-1}\mathbf{W}^{\top}\mathbf{Q}_2^{\top}\right)\\
    \overset{\triangle}{=}&\mathsf{Tr}\left(\mathbf{P}_1^{-1}\mathbf{Q}_1\right) + \mathsf{Rest}
    \end{aligned}
\end{equation*}
where $\mathsf{Tr}\left(\mathbf{P}_1^{-1}\mathbf{Q}_1\right)$ does not depend on $\mathbf{W}$ and using the cyclic property of trace, the rest terms can be further expressed as 
\begin{equation*}
    \begin{aligned}
        \mathsf{Rest}&=\mathsf{Tr}\left(\mathbf{D}^{-1}\mathbf{W}^{\top}\mathbf{\Xi}\mathbf{W}\right)\\
    & = \mathsf{Tr}\left((\mathbf{W}^{\top}(\mathbf{P}_3-\mathbf{\Upsilon})\mathbf{W})^{-1}\mathbf{W}^{\top}(\mathbf{Q}_3 -\mathbf{\Xi})\mathbf{W}\right),
    \end{aligned}
\end{equation*}
with $\mathbf{\Xi}=\mathbf{P}_2^{\top}\mathbf{P}_1^{-1}\mathbf{Q}_1\mathbf{P}_1^{-1}\mathbf{P}_2 + \mathbf{P}_2^{\top}\mathbf{P}_1^{-1}\mathbf{Q}_2 - \mathbf{Q}_2^{\top}\mathbf{P}_1^{-1}\mathbf{P}_2$. This completes the proof.

\bibliographystyle{IEEEtran}
\bibliography{Ref}

\end{document}